\documentclass[letterpaper,twocolumn,10pt]{article}

\usepackage{usenix2019_v3}
\usepackage[utf8]{inputenc}

\usepackage[square,numbers,compress]{natbib}
\usepackage{multirow}
\usepackage{bm}
\usepackage{algorithmicx}
\usepackage{algpseudocode}
\usepackage{enumitem}
\usepackage{color}
\usepackage{amsmath,amsfonts,mathtools,nccmath,amsthm,commath,amssymb}
\usepackage{caption}
\usepackage{subcaption}
\usepackage{xurl}
\usepackage{listings}
\usepackage{pifont}
\usepackage{hyphenat}
\usepackage{textcomp}
\usepackage[ruled,vlined,linesnumbered]{algorithm2e}
\usepackage{makecell}

\let\oldnl\nl
\newcommand{\nonl}{\renewcommand{\nl}{\let\nl\oldnl}}

\newcommand \BROKEN {\textcolor{red}{\ding{55}}}
\newcommand \SECURE {\textcolor{blue}{\ding{51}}}

\newcommand \proposed {\textit{FuncTeller}}

\begin{document}

\title{\proposed: How Well Does eFPGA Hide Functionality?}

\author{
{\rm Zhaokun Han{$^\dagger$}, Mohammed Shayan{$^\ast$},  Aneesh Dixit{$^\dagger$}, Mustafa Shihab{$^\ast$},} \\
{\rm Yiorgos Makris{$^\ast$}, and Jeyavijayan (JV) Rajendran{$^\dagger$}} \\
{$^\dagger$}Texas A\&M University, {$^\ast$}The University of Texas at Dallas  \\        
{$^\dagger$}\texttt{\{hzhk0618,aneeshdixit,jeyavijayan\}@tamu.edu}, \\
{$^\ast$}\texttt{\{shayan.mohammed,mustafa.shihab,yiorgos.makris\}@utdallas.edu}}

\maketitle

\begin{abstract}

Hardware intellectual property (IP) piracy is an emerging threat to the global supply chain. Correspondingly, various countermeasures aim to protect hardware IPs, such as logic locking, camouflaging, and split manufacturing. However, these countermeasures cannot always guarantee IP security. A malicious attacker can access the layout/netlist of the hardware IP protected by these countermeasures and further retrieve the design. To eliminate/bypass these vulnerabilities, a recent approach redacts the design's IP to an embedded field-programmable gate array (eFPGA), disabling the attacker's access to the layout/netlist. eFPGAs can be programmed with arbitrary functionality. Without the bitstream, the attacker cannot recover the functionality of the protected IP. Consequently, state-of-the-art attacks are inapplicable to pirate the redacted hardware IP. In this paper, we challenge the assumed security of eFPGA-based redaction. We present an attack to retrieve the hardware IP with only black-box access to a programmed eFPGA. We observe the effect of modern electronic design automation (EDA) tools on practical hardware circuits and leverage the observation to guide our attack. Thus, our proposed method \proposed~selects minterms to query, recovering the circuit function within a reasonable time. We demonstrate the effectiveness and efficiency of \proposed~on multiple circuits, including academic benchmark circuits, Stanford MIPS processor, IBEX processor, Common Evaluation Platform GPS, and Cybersecurity Awareness Worldwide competition circuits. Our results show that \proposed~achieves an average accuracy greater than $85$\% over these tested circuits retrieving the design's functionality.

\end{abstract}

\section{Introduction}
\label{sec:introduction}

\subsection{Hardware 
IP: Threats and Defenses} \label{sec:hardware_security}

The monetary loss associated with intellectual property (IP) theft is comparable to the amount of US exports to Asia~\cite{ipcom2013}.
This ongoing theft results in a loss of revenue for the IP developers and diminishes incentives for investment in research and development.
Many leading semiconductor companies outsource design and manufacturing while owning the hardware IP.
In 2020, such companies constituted 33\% of the entire semiconductor market~\cite{fabless}. Consequently, there is an economic incentive for reverse engineering, unauthorized usage, and overproduction of hardware IPs.  
For example, according to a report from the US Department of Justice in 2018, 
the worldwide supply for dynamic random access memory (DRAM) is worth nearly \$100 billion, and Micron controls $20$-$25\%$ of the DRAM industry; however,
IP theft in Micron
caused an estimated loss of \$8.75 billion~\cite{doj2018}. These deleterious consequences underline the need for countermeasures against IP theft in the semiconductor industry.

Semiconductor fabrication facilities are concentrated in countries prone to geopolitical tensions and conflicts~\cite{bcgreport}. 
Except for Intel, IBM, and Samsung, nearly all chip fabrication is carried out in such foreign territories.
Consequently, the supply chain of the semiconductor industry is vulnerable to threats~\cite{Tuomi2009TheFO}.
Since untrusted foundries and testing facilities have access to the hardware IP, rogue employees from these entities
may attempt to pirate the IP. 
Additionally, an untrusted end-user can purchase a chip from the market and use reverse engineering to extract the hardware IP netlist.

\begin{table*}[thb!]
    \centering
    \caption{The relationship between various hardware security threats and countermeasures. A \SECURE/\BROKEN~denotes the countermeasure is secure against/vulnerable to the threat, and ``N.A.'' denotes the threat is not applicable to the countermeasure.}
    \resizebox{0.995\textwidth}{!}{
    \begin{tabular}{|l||c|c|c|c|}
    \hline
    \multirow{2}{*}{{\bf Threat} } 
    & \multicolumn{4}{c|}{\bf Countermeasure} \\ \cline{2-5}
    & {\bf Logic locking~\cite{epic, AntiSAT2}} & {\bf Camouflaging~\cite{camoStdCell}} & {\bf Split manufacturing~\cite{jarvis2007split, ImesonSEC13}} & {\bf FPGAs~\cite{bandari2021iccad,bhandari2021not,mohan2021hardware,DisableBitstreamXilinx,AgainstDPA,SummaryFPGAProtect}}  \\ \hline\hline
    {\bf IP piracy~\cite{CADpiracy}}
    & \SECURE & \SECURE & \SECURE & \SECURE \\ \hline
    {\bf Reverse engineering~\cite{torrance2009state}}
    & \SECURE & \SECURE & \SECURE & \SECURE \\ \hline
    {\bf Algorithmic attacks~\cite{pramod2015sat,el2019sat,wang2016cat, li2019attacking,MengliProvablySecureCamo,pramod2019funct}}
    & \BROKEN & \BROKEN & \BROKEN & \SECURE \\ \hline
    {\bf Bitstream extraction~\cite{2012bitstream,bitstream2018,yang2004scan, szefer2019survey, koeune2005tutorial}}
    & N.A. & N.A. & N.A. & \SECURE \\ \hline
    \end{tabular}
    }
    \label{tab:general_relationship_threats_countermeasures}
\end{table*}

Numerous countermeasures have been proposed to prevent hardware IP theft: logic locking, camouflaging, and split manufacturing
are three prominent examples~\cite{epic, AntiSAT2, camoStdCell, jarvis2007split,ImesonSEC13}. 
Logic locking protects the design by adding extra logic controlled by additional key inputs~\cite{epic, AntiSAT2}; only the correct key restores the correct functionality. 
Logic locking can defend against reverse engineering as it {\it obscures} the IP.
Camouflaging deters reverse engineering attacks from end-users by designing a chip that looks alike but performs different functionalities, which is known only to the owner~\cite{camoStdCell}. 
Split manufacturing protects the IP by splitting the manufacturing process among different foundries, thereby preventing a single foundry from obtaining the complete design information~\cite{jarvis2007split,ImesonSEC13}.
While these approaches increase trust in the supply chain, they are still vulnerable to several attacks.

The existing IP protection techniques become vulnerable when the attacker has oracle access, i.e., a functional chip sold on the market. 
For instance, some defenses are vulnerable to input-output (I/O) query-based attacks~\cite{pramod2015sat,el2019sat}, where selected input patterns are applied, and the output differences between the locked design and the oracle are observed to extract the key.
In addition, the attacker can analyze the structural traces in the locked netlist and recover the original design;
these attacks are referred to as structural attacks~\cite{MengliProvablySecureCamo,pramod2019funct,doesLL}. 
Collectively, I/O and structural attacks render most logic locking, camouflaging, and split manufacturing techniques vulnerable, as listed in Table~\ref{tab:general_relationship_threats_countermeasures}.

\subsection{FPGA-based IP Protection}
\label{sec:intro_hardware redaction}
Given the weakness of existing IP protection techniques, researchers have proposed to redact the complete design information from the untrusted supply chain using a field-programmable gate array (FPGA). 
An FPGA is a reprogrammable integrated circuit that can be programmed to have an arbitrary functionality by uploading a bitstream to the FPGA. 
In contrast, traditional application-specific integrated circuits (ASICs) are designed to perform a specified function and cannot be reconfigured. 
{While ASICs comprise multiple gates derived from a specified technology library}, FPGAs consist of multiple programmable logic blocks. These logic blocks, in turn, consist of multiple lookup tables (LUTs), which are programmable based on the desired functions.
Therefore, an FPGA netlist appears as a cluster
of LUTs and does not provide any meaningful information unless programmed with a bitstream.
Thus, compared to the conventional integrated circuit (IC) design model, the hardware IP is the bitstream rather than the FPGA platform itself.

Xilinx FPGAs can protect the IP at the software level by encrypting the bitstream and preventing the attacker from using the \textit{readback} functionality to extract the bitstream~\cite{2012bitstream,bitstream2018,DisableBitstreamXilinx}. 
Recent attacks attempting to recover the bitstream have been published~\cite{2012bitstream,bitstream2018,yang2004scan, szefer2019survey, koeune2005tutorial}---however, advanced bitstream protection mechanisms 
thwart
these attacks~\cite{DisableBitstreamXilinx,SummaryFPGAProtect,AgainstDPA}. 
These protections safeguard the bitstream at the software and hardware levels.
Thus, due to their security and versatility (as shown in Table~\ref{tab:general_relationship_threats_countermeasures}),  FPGA-based countermeasures have gained interest in academia and industry~\cite{CmuSlidesDarpa,sahara,intelnx5,bandari2021iccad,bhandari2021not,mohan2021hardware}.

FPGAs can also be embedded into ASIC, referred
to as embedded FPGAs (eFPGAs).
A design implemented on an ASIC with an eFPGA offers
the reconfigurability of FPGAs while mostly retaining the 
power, performance, and area (PPA) cost-effective
benefits of ASICs.
eFPGAs have also piqued interest due to their security capabilities. 
Particularly, it has led to the development of an IP piracy countermeasure known as eFPGA-based hardware redaction.
This IP protection has captured significant interest~\cite{bandari2021iccad,bhandari2021not,mohan2021hardware}.
Obfuscated Manufacturing for GPS (OMG)~\cite{CmuSlidesDarpa} and Structured Array Hardware for Automatically Realized Applications (SAHARA)~\cite{sahara} are two notable projects supported by the Defense Advanced Research Projects Agency (DARPA) utilizing this technology.
The recent Cybersecurity Awareness Worldwide (CSAW) 2021 competition supported by Siemens \textit{et al.} also included designs redacted by eFPGAs~\cite{csaw2021}.
Importantly, the US military has selected Intel’s structured ASIC technology, a hybrid of FPGA and ASIC, to protect hardware IPs~\cite{intelnx5,sahara}.

\subsection{Our Goals and Contributions}
In this work, we ask a question: {\it How secure are eFPGAs?} We answer it by  attempting to recover an IP implemented on an eFPGA with only I/O (oracle) access.  
To this end,
we develop a heuristic attack, \proposed, overcoming the following challenges of recovering hardware IPs 
on eFPGAs:

\begin{enumerate}[leftmargin=*]
\item[1.] 
The size of the search space for an attacker is $2^n$, where $n$ is the number of inputs of the redacted design implemented on the eFPGA. Many practical hardware designs have a large number of inputs,
e.g., the IBEX processor has $1,386$ inputs~\cite{woods2014ibex}. 
This renders brute-force search impractical and forces the attacker
to smartly choose input patterns for retrieving the redacted design (Section~\ref{sec:previous_attacks}).

\item[2.]
A heuristic algorithm may predict approximate
functionality, but it sacrifices accuracy for efficiency.
Thus any practical attack must ensure accuracy scaling, particularly for hardware IPs with a large number of inputs (Section~\ref{sec:previous_attacks}).

\item[3.]
eFPGAs have a generalized structure independent of the implemented 
design, and
the number of I/O pins is design-agnostic.
Generally, some output pins are unused by design.
Such unused output pins are driven to a constant $0$/$1$.
For example, the IBEX processor~\cite{woods2014ibex} implemented on an FPGA with $18$~K gates has $117$ outputs with constant $0$/$1$ functionality. 
Identifying these corner cases is crucial for reducing the execution
time of a successful attack. 
\end{enumerate}

To overcome these challenges, we leverage several properties of hardware designs to develop a practical and effective attack, \proposed. The salient features of \proposed~are: 
Firstly, \proposed~predicts the Boolean function of a hardware IP by querying a small number of input patterns within the entire search space of size $2^n$. The prediction is made by carefully selecting the input patterns, leveraging the fact that the input patterns with similar behavior are clustered together, and the distance between them is small (Section~\ref{sec:proposed_expanding}). 
Secondly, \proposed~parameterizes the prediction algorithm, enabling the user to find the appropriate trade-off between accuracy and efficiency. 
These parameters are empirically derived to optimize the \proposed~prediction (Section~\ref{sec:proposed_expanding}). 
Thirdly, \proposed~predicts each output independently. Thus, it allows the attack to be executed parallelly on each output, thereby speeding up the overall prediction (Section~\ref{sec:proposed_single_cone}).
Finally, \proposed~identifies special 
cases, such
as constant logic $0/1$ outputs, allowing it to reduce prediction time and improve accuracy
(Section~\ref{sec:recover_single_cone}).

The paper's contributions can be summarized as follows:
\begin{enumerate}[leftmargin=*]
  \item We present a heuristic solution, \proposed, to extract redacted hardware IP with high 
  accuracy, for \proposed~achieves an accuracy of $91\%$ on the IBEX processor~\cite{woods2014ibex} (Section~\ref{sec:simu}).
  \item \proposed~can 
  achieve 
  $1.22\times$ more accuracy than
  the existing state-of-the-art attack~\cite{chen2020circuit}
  for predicting a black-box circuit's functionality (Section~\ref{sec:simu}). 
\item 
We evaluate the performance of \proposed~on various open-sourced and widely-used circuits,
including
seven
academic benchmarks (ISCAS'85~\cite{iscas85} and ITC'99~\cite{itc99}) and three industrial circuits, such as Stanford MIPS processor~\cite{hennessy1981mips} ($4$~K gates),
IBEX processor~\cite{woods2014ibex} ($18$~K gates),
and Common Evaluation Platform 
(CEP)
GPS circuit~\cite{darpa_cep} ($213$~K gates) (Section~\ref{sec:attacks_on_practical_circuits}).   
\item  We present an analysis of the trade-off between accuracy and efficiency while running \proposed.
  {In a case study on IBEX~\cite{woods2014ibex}, the trade-off curve describes the variation in accuracy from $81.2\%$ to $88.5\%$.}
  This analysis allows the attacker to ascertain the estimated accuracy for a given time 
  limit
  before running \proposed~(Section~\ref{sec:trade-offs}). 
  \item 
  {The paper discusses potential countermeasures to defeat \proposed, to motivate the development of better countermeasures to safeguard the hardware IP
  (Section~\ref{sec:potential_countermeasures}).}
\end{enumerate}

In addition to the listed contributions,
we plan to open-source our attack tool.

\section{Background and Prior Work}
\label{sec:back}

This section describes an intellectual property (IP) piracy countermeasure using embedded field-programmable gate arrays (eFPGAs) and discusses why multiple attempts to circumvent the countermeasure have failed. Finally, we detail hardware design principles from the field of logic synthesis. These principles will form the backbone of our attack.

\subsection{eFPGA-based Hardware Redaction}
\label{sec:efpga_background}

\begin{figure}[bt!]
    \centering
    \includegraphics[clip, trim=1.2cm 0.9cm 0.7cm 1cm,
    width=0.48\textwidth]{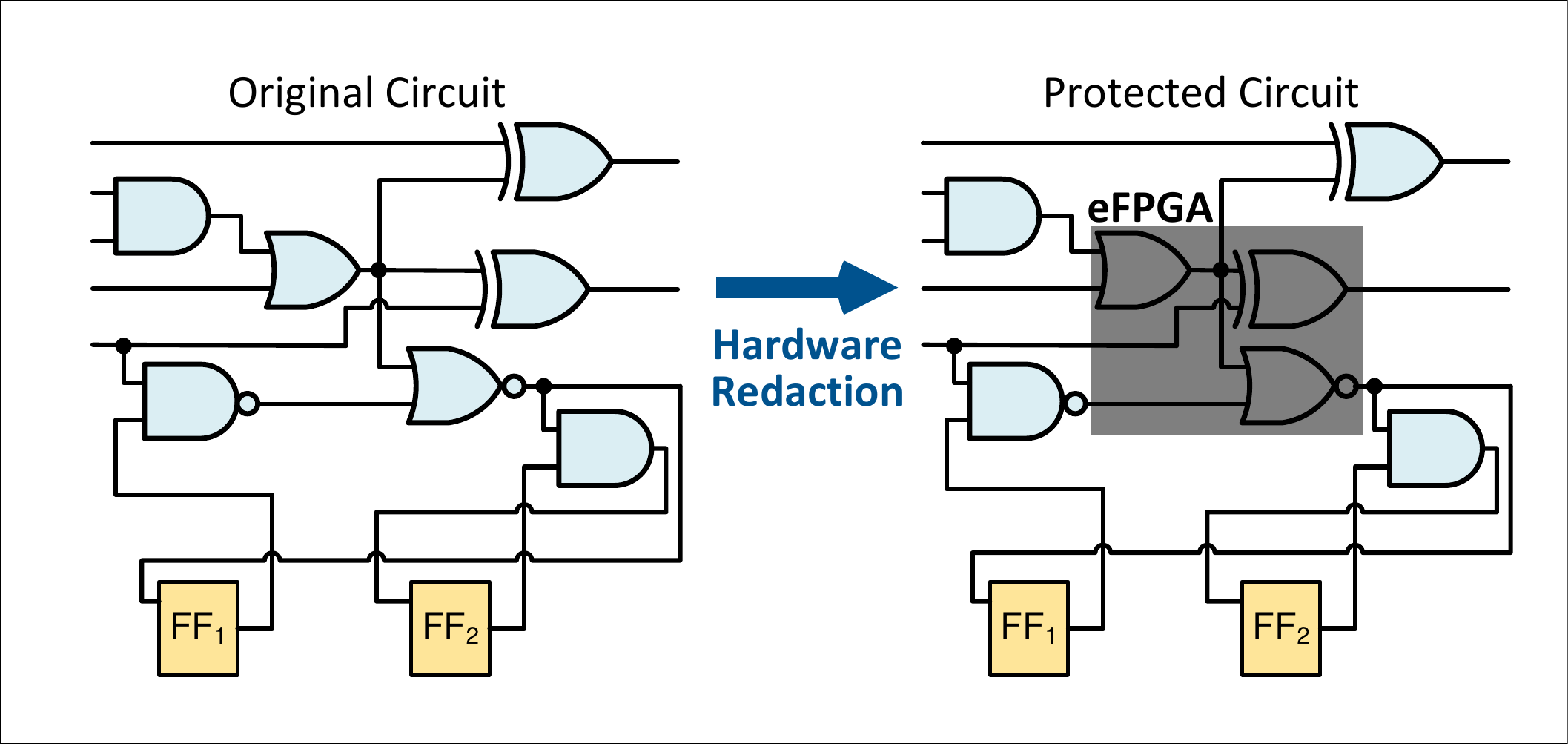}
    \caption{Design netlists before and after hardware redaction. Gates and wires inside the black box are eFPGA-redacted.}
    \label{fig:hw_redaction_protection}
\end{figure}

As described in Section~\ref{sec:intro_hardware redaction}, a field-programmable gate array (FPGA) netlist does not provide valuable information to an attacker. This fact is utilized to design a special class of circuits wherein a traditional application-specific integrated circuit (ASIC) is embedded with an FPGA. This eFPGA can implement a part of the hardware IP. Thus, \textit{``hiding''} or redacting the said part of the circuit. Post-manufacturing, the eFPGA is loaded with the bitstream.
Due to the lookup table (LUT) based structure of FPGAs, the netlist does not reveal the hardware IP that it has been programmed with.
Using an eFPGA to \textit{``hide''} the IP is termed as eFPGA-based hardware redaction~\cite{bandari2021iccad,bhandari2021not}, as represented in Figure~\ref{fig:hw_redaction_protection}.

\subsection{Prior Work}
\label{sec:previous_attacks}
All digital circuits implement a Boolean function. Thereby, the problem of extracting hardware IP from a  black-box design is similar to the problem of learning the Boolean function implemented by the design. Thus, we can observe the responses of black-box design to chosen queries and use the results to recover the Boolean function or its approximation.

Multiple works follow the same principle and provide solutions for learning a Boolean function~\cite{dnflearning,incrlearning}.
However, these algorithms are borrowed from the field of circuit design and are designed for traditional ASICs requiring a netlist.
As discussed in Section~\ref{sec:efpga_background}, a meaningful netlist with design-specific information is not accessible for eFPGA-redacted hardware.  Without the netlist, Angluin \textit{et al.}~\cite{boundedlearning} show that these learning algorithms require querying all $2^n$ input patterns, where $n$ is the number of inputs, rendering these algorithms inefficient for practical scenarios.

A Satisfiability (SAT) based approach~\cite{pramod2015sat} can successfully retrieve IP
for ASIC designs, even if the circuit is protected by state-of-the-art IP protection mechanisms, by solving conjunctive normal forms (CNFs) with a SAT solver.
However, this approach does not apply to FPGA designs because FPGAs consist of programmable LUTs.
LUTs with large input sizes result in a computational complexity exponential to the number of inputs.
Thus, applying 
SAT solvers
to deduce the designs implemented on FPGAs is computationally infeasible.
A recent work on deobfuscating eFPGAs, Rezaei \textit{et al.}~\cite{rezaei2022evaluating}, proposes to apply a SAT-based approach that simplifies the CNF conditions (eliminating structural loops) by leveraging the usage information/parameters of the FPGA (such as the input size and the number of used LUTs)~\cite{rezaei2022evaluating}. 
However, these parameters (the input size and the number of used LUTs) are not accessible to the attacker under our  threat model,  where the attacker has only black-box access.

Generic logic regression-based approaches~\cite{ruczinski2003logic} attempt to recreate a Boolean function by fitting a model on a collection of data points. These data points are obtained by querying the oracle with input patterns and observing outputs.
However, this approach does not consider properties specific to practical hardware circuits and operates on the entire Boolean space. Further, it results in low efficiency and lack of scalability to larger circuits,
as shown in Section~\ref{sec:previous_work_limitation}.

Chen \textit{et al.}~\cite{chen2020circuit} propose a circuit learning approach based on Boolean regression. To the best of our knowledge, this presents the state-of-the-art attack to retrieve black-box Boolean functions. Chen \textit{et al.}~\cite{chen2020circuit} overcome the severe drawback of generic logic regression methods by considering properties specific to hardware circuits. 
However, this approach maps the outputs to a library of Boolean functions to speed up the algorithm, and this speed benefit may not be universally applicable to all circuits. 
Thus, its practicality is limited.

\color{black}
\subsection{Limitations of Existing Approaches}
\label{sec:previous_work_limitation}
The logic regression techniques mentioned in Section~\ref{sec:previous_attacks} face issues while scaling up to large circuits.
More importantly, these logic regression techniques have low accuracy. 
For example, the state-of-the-art tool from Chen \textit{et al.}~\cite{chen2020circuit} can recover IBEX with an accuracy of only $56.68$\%. 
Further, we observed that some IPs, such as 
Common Evaluation Platform (CEP)
GPS~\cite{darpa_cep}, cannot be recovered even after
three days of run-time.
These techniques choose
queries randomly without adapting them to the  underlying
practical hardware design properties. This hardware-agnostic approach results in low accuracy and efficiency and is a deficiency for all traditional logic regression techniques, including Chen \textit{et al.}~\cite{chen2020circuit}.

\proposed~overcomes these drawbacks by focusing on: (i) specificity towards practical hardware
designs, (ii) scalability, (iii) accuracy, and (iv) usability.
\proposed~specifically targets practical hardware IPs in contrast to previous works, which are mainly extensions of theoretical solutions. 
As a result, \proposed~significantly improves accuracy and can scale to larger designs. Additionally, \proposed's heuristic nature provides 
users flexibility
through user-defined parameters and allows them to tune the attack to different settings.

\subsection{Logic Synthesis}
\label{sec:logic_synthesis}

As mentioned in Section~\ref{sec:previous_work_limitation}, \proposed~considers hardware-specific properties which enable it to outperform previous attacks. These properties are rooted in fundamental logic synthesis principles, which are outlined in this subsection.

Consider a Boolean function $f$ with single output and multiple inputs,
$a$, $b$, and $c$. The Boolean function is written as
$f = abc + ab\overline{c} $.
This representation of the function is referred to as the {\it sum of products (SOP)}, and every individual product is referred to as a minterm. 
Here, each product can be regarded as an {\it ON-set  minterm} if it returns $f=1$ when it is an input; otherwise, an {\it OFF-set minterm} if the result is $f=0$.
Note that there are $2^n$ minterms for an $n$-input function, and each minterm is either ON-set or OFF-set.

For some cases, multiple ON-set minterms can be compressed/merged into an {\it implicant}.
The implicant is represented by elements from $\{0, 1, \scalebox{1.3}{-} \}$, where ``$\scalebox{1.3}{-}$'' is a 
{\it don’t care}
covering both $0$ and $1$ cases.
For example, $I_1=ab\,\scalebox{1.3}{-}$ is an implicant with one don’t care bit. 
$I_1$ covers two minterms: $abc$ and $ab\overline{c}$.
Therefore, the Boolean function of $f$ can be simplified as $f = ab$.

\begin{figure}[htb!]
    \centering
     \begin{subfigure}[b]{0.22\textwidth}
         \centering
         \includegraphics[clip, trim=1cm 0.8cm 7cm 0.8cm,
         width=1\textwidth]{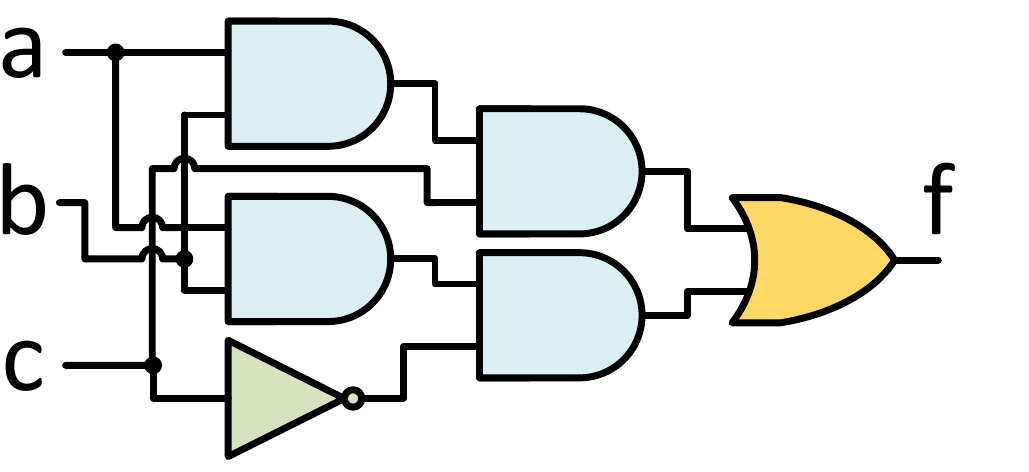} 
         \caption{}
     \end{subfigure}
    \begin{subfigure}[b]{0.11\textwidth}
         \centering
         \includegraphics[clip, trim=12cm 0.8cm 1cm 1cm,
         width=1\textwidth]{Figure/sop.pdf} 
         \caption{}
    \end{subfigure}
    \caption{The circuits of Boolean function $f=abc + ab\overline{c}$ (a)  before and (b) after synthesis.}
    \label{fig:PIT_simplification}
\end{figure}

Most logic synthesis algorithms aim to reduce the cost of logic implementation by removing redundant implicants.
Implicants uncovered by other implicants are called \textit{prime implicants (PIs)}.
A set of PIs containing all ON-set minterms forms a {\it prime implicant table (PIT)}.
For most practical designs, PIs are distributed close to each other. PIs are rarely far apart, according to Han \textit{et al.}~\cite{doesLL}: this is 
an attribute of the Boolean functions of these designs.
In the modern hardware design cycle, powerful commercial tools, such as \textit{Synopsys Design Compiler}~\cite{designcompiler}, \textit{Cadence Genus}~\cite{genus}, and \textit{Siemens Precision RTL}~\cite{precisionRTL}, utilize this attribute of Boolean functions while performing synthesis processes to optimize the circuit, simplify the logic, and reduce hardware cost.
For example, consider the Boolean function of $f = abc+ab\overline{c}$. If each of the function's minterm is implemented individually (without any optimization), the function can be realized using six logic gates, as shown in Figure~\ref{fig:PIT_simplification}(a).
However, the same function can be simplified/synthesized to  $f=I_1=ab\,\scalebox{1.3}{-}$, 
so the circuit can be realized with one logic gate, as shown in Figure~\ref{fig:PIT_simplification}(b). This type of optimization is a hallmark of all synthesis tools that aggressively minimize the number of PIs in order to minimize the number of gates, which in turn minimizes the circuit's area and power. \proposed~exploits the fact that most practical designs have this property (PIs are distributed close to each other) so that synthesis tools can utilize it
and simplify PIT.

\section{Threat Model} \label{sec:threat_model}

We consider a threat model consistent with previous works on
embedded field-programmable gate array (eFPGA) based
hardware redaction~\cite{bandari2021iccad},
attacks on eFPGA~\cite{chowdhury2022predictive, rezaei2022evaluating},
and the Cybersecurity Awareness Worldwide competition~\cite{csaw2021}.

\noindent
{\bf Attacker's Locations:} The attacker could be a collusion of an
untrusted foundry/testing facility and an
untrusted end-user. The attacker in the foundry/testing facility has access to a chip netlist with the unprogrammed eFPGA. The attacker can procure a functional chip from the market and can isolate eFPGA by analyzing the chip netlist and identifying the 
scan chains
connected to the eFPGA using 
reverse engineering~\cite{torrance2009state}.
Note that the purchased functional chip is the application-specific integrated circuit (ASIC) integrated with the configured (loaded with a certain bitstream) eFPGA.

\noindent{\bf Attacker's Capabilities:}
In this threat model, the attacker has the following capabilities:

\begin{itemize}[leftmargin=*]
    \item 
    The attacker can isolate the eFPGA from the rest of the design by accessing dedicated scan chains of eFPGA, a feature commonly supported by eFPGA vendors~\cite{chowdhury2022predictive}. Alternatively, the attacker can perform a probing attack to isolate eFPGA by locating and accessing the ASIC's internal signals to control/observe eFPGA's inputs/outputs~\cite{wang2017probing,chowdhury2022predictive}.
    
    \item An attacker is unable to
    retrieve the hardware intellectual property (IP) by performing a side-channel attack or bitstream extraction. Over the years, multiple schemes have been proposed to mitigate these attacks~\cite{yang2004scan, szefer2019survey, koeune2005tutorial,SummaryFPGAProtect,AgainstDPA}.
    
    \item After isolating the eFPGA, the attacker can enable scan-chain access by stripping scan-chain protections~\cite{DaRolt2011scanattackandcountermeaures,ali2015novel,alrahis2019scansat,limaye2020dynunlock}. 
    Further, the attacker can access input-output (I/O) pins through scan chains to query the hardware IP design.
    Appendix~\ref{sec:appendix_scan_chain} provides more details on the mechanisms to enable or unlock scan-chain access,
    including attacks
    on scan-chain protections~\cite{DaRolt2011scanattackandcountermeaures, ali2015novel, alrahis2019scansat, limaye2020dynunlock}.
\end{itemize}

\noindent
\textbf{Attacker's Goal:}
The attacker aims to accurately and efficiently recover the hardware IP's functionality by only querying the 
black-box combinational part of the hardware design.

\section{\proposed: Technical Approach}
\label{sec:proposed}

In this section, we present our attack, \proposed, to recover a design implemented on an eFPGA
under the threat model described 
in Section~\ref{sec:threat_model}.
Recall that logic synthesis on practical circuits optimizes 
power, performance, and area (PPA) by minimizing the
number  
of prime implicants
(PIs)
and literals
in the circuit's prime implicant table (PIT).
As a consequence, logic synthesis clusters
ON-set
minterms
into several PIs.
This behavior is a consistent feature of practical hardware designs and has the following two implications:

\begin{itemize}[leftmargin=*]
\item 
A single PI covers multiple ON-set minterms that are consistent with the literals in the PI representation.
We use this property to predict each PI by expanding
from a discovered ON-set minterm (seed): 
attempting to replace each literal with a don't care
and heuristically verifying each replacement.

\item
The distance between any two PIs in a PIT is usually much smaller than the input size. We use this property to reduce the search space when updating the predicted PIT with the new PI. The generation of new PI requires the discovery of the next ON-set minterm. Therefore, we limit the search space for the next ON-set minterm to being in close proximity to the current PIs.
\end{itemize}

\begin{figure}[htb!]
    \centering
     \begin{subfigure}[b]{0.25\textwidth}
         \centering
         \includegraphics[clip, trim=0.3cm 0.8cm 11cm 0.8cm,
         width=1\textwidth]{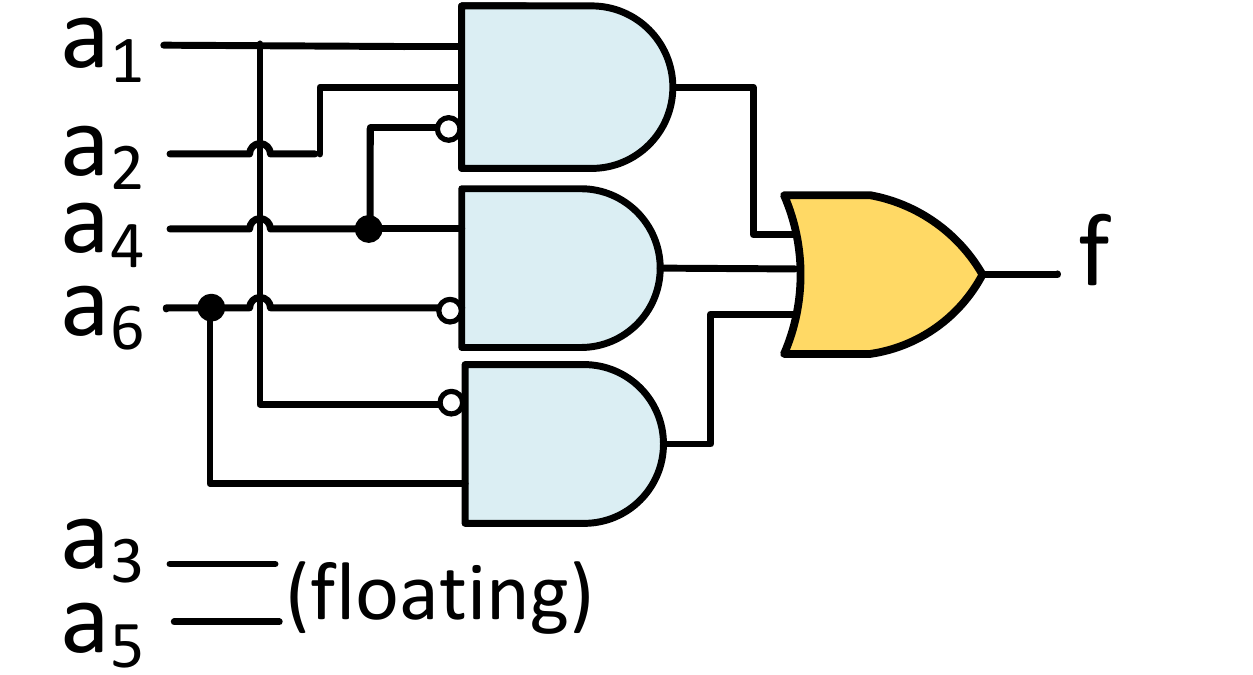} 
         \caption{}
     \end{subfigure}
    \begin{subfigure}[b]{0.19\textwidth}
         \centering
         \includegraphics[clip, trim=13.5cm 0.4cm 0.75cm 0.8cm,
         width=1\textwidth]{Figure/original_circuit_structure.pdf}
         \caption{}
     \end{subfigure}
    \caption{An example circuit: (a) the netlist and (b) the PIT.}
    \label{fig:original_netlist_and_pit}
\end{figure}

For the remainder of Section~\ref{sec:proposed}, we show \proposed~exploits 
these
implications.
Further, we describe the mechanism of \proposed, using 
the example circuit in
Figure~\ref{fig:original_netlist_and_pit}.
Consider a circuit cone\footnote{A multi-output circuit could be collapsed into several $1$-output circuits. Such a $1$-output circuit is referred to as a cone.} with $n$ inputs ($a_1,a_2,$ $\cdots, a_n$) and one output;
the circuit implements the Boolean function $f$ 
with $n=6$,
as shown in Figure~\ref{fig:original_netlist_and_pit}(a). 
The Boolean function is  
$f(a_1,a_2,\cdots,a_6) = a_1a_2\overline{a_4} + a_4\overline{a_6} + \overline{a_1}a_6$,
and its PIT is shown in Figure~\ref{fig:original_netlist_and_pit}(b). 
Note that the expression of PIT for a certain Boolean function is non-canonical.\footnote{A circuit representation, which is not unique, is considered non-canonical. Here, there may exist multiple valid PITs for the same output.}

\subsection{Initialization} \label{sec:proposed_initialization}
\proposed~initializes by finding an ON-set minterm $m_0$. 
{We search for $m_0$}
by querying the oracle with random minterms,
and it ends upon finding an ON-set minterm within $r$ queries ($r\in \mathbb{N}^{+}$). 
If all the $r$ queries are OFF-set minterms, then \proposed~regards the output's function as constant $0$ and terminates out of the purpose of efficiency.

\subsection{Expanding from ON-set Minterm to PI}
\label{sec:proposed_expanding}
\color{black}

After finding an ON-set minterm $m_0$, 
\proposed~predicts the PI 
containing
$m_0$. This is referred to as ``the expansion of the ON-set minterm $m_0$ to a predicted PI $PI_{pred}$''. 
Algorithm~\ref{alg:minterm2PI} describes the corresponding routine.
We first set a variable string ``$cube$'' equal to the binary pattern of the ON-set minterm $m_0$. We update each literal of $cube$ to form predicted PI $PI_{pred}$.

For each ($i^{th}$) literal of the cube, we recursively construct a new group of minterms 
$\{m_1\}$ by inverting the $i^{th}$ $cube$ literal and replacing the don't cares (if any) with binary ``1''/``0''. After constructing each minterm 
$m_1$,
we query it
using the oracle $\mathcal{O}$. 
If the query shows that
$m_1$
is in the OFF-set,
we
stop generating new minterm 
$m_1$
corresponding to the $i^{th}$ literal  
inversion. Instead, we keep the $i^{th}$ literal as in $m_0$ and move to the next literal.
On the other hand, if all the minterms formed by inverting the $i^{th}$ $cube$ literal belong to 
the ON-set,
then the $i^{th}$ $cube$ literal is updated to a don't care ``$\scalebox{1.3}{-}$'' and then we move to the next literal. 
Once it has been through all literals in $m_0$,
the algorithm returns the updated $cube$ as $PI_{pred}$.

During the PI prediction,
the number of
minterms to be 
queried
is exponential to the number of don't cares in the $cube$.
Algorithm~\ref{alg:minterm2PI} uses a heuristic approach to avoid querying all 
covered
minterms
and can be inaccurate while predicting. 
The probability of prediction error depends on the number of minterms queried during the  prediction routine. 
To control the error while keeping \proposed~efficient,
we introduce a ``linear limitation parameter $p$.''
When updating each literal, the maximum number of queries is capped at $p$ times the number of don’t cares. 
Thus,
the PI prediction for an $n$-input function requires 
querying at most $p \times n^2$ minterms. 
Thus, the algorithm takes time complexity $O(p \times n^2) = O(n^2)$. We further discuss the performance improvement of Algorithm~\ref{alg:minterm2PI} in Appendix~\ref{sec:proposed_scable_expansion} and Appendix~\ref{sec:appendix_parameters_accuracy}.

\begin{algorithm}[tb!]
\caption{Expansion of an ON-set minterm to PI}
\label{alg:minterm2PI}

\small

\SetKwFunction{FMain}{\text{expand\_minterm\_to\_PI}}
\SetKwFunction{FUpdateCube}{update\_cube}

\SetKwProg{Fn}{Function}{:}{}

\SetKwProg{FORwithnoend}{for}{ do}{}
\SetKwProg{IFwithnoend}{if}{ then}{}

\SetAlgoLined
\KwIn{Oracle $\mathcal{O}$, output index $w$, ON-set minterm $m_0$,
{linear limitation parameter $p$}
}
\KwOut{Predicted PI $PI_{pred}$}

\Fn{ \FMain {$\mathcal{O}$, $w$, $m_0$, $p$}}{
$cube$ {$:=$}  $m_0$ \hfill {$\rhd$Initialize $cube$} \\

\FORwithnoend{$index \in \{1,2, \cdots, len(m_0)\}$}{
$num\_dc$ {$:=$}  $cube.count($``$\scalebox{1.35}{-}$''$)$\\
$iter\_limit$ {$:=$}  $min\{2^{num\_dc}, p\times num\_dc \}$ \\ 

$cube := $ \FUpdateCube($\mathcal{O}$, $w$, $cube$, $index$, $iter\_limit$) \\

}
$PI_{pred}$ {$:=$} $cube$ \\
}
{\bf {return}} $PI_{pred}$  \\

\Fn{\FUpdateCube{$\mathcal{O}$, $w$, $cube$, $index$, $iter\_limit$}}{
$flag$ {$:=$}  $1$\\
\FORwithnoend{$i \in \{1,2,\cdots, iter\_limit \}$}{
    $m_1$
    {$:=$}  pick\_random\_minterm$(cube, index) $ \\
    $response$ {$:=$}  $ \mathcal{O}.\text{query\_oracle}(m_1, w)$ \\
    \IFwithnoend{$response == 0$}{
        $flag$ {$:=$}  $0$ \hfill{$\rhd$Fail to replace with don't care}  \\ 
        {\bf break} \\
    }
}
\IFwithnoend{$flag == 1$}{
$cube := replace\_with\_dc\_bit(cube, index)$ \\
}
}
{\bf {return}} $cube$

\end{algorithm}

\begin{figure}[htb!]
    \centering
    \includegraphics[width=0.43\textwidth]{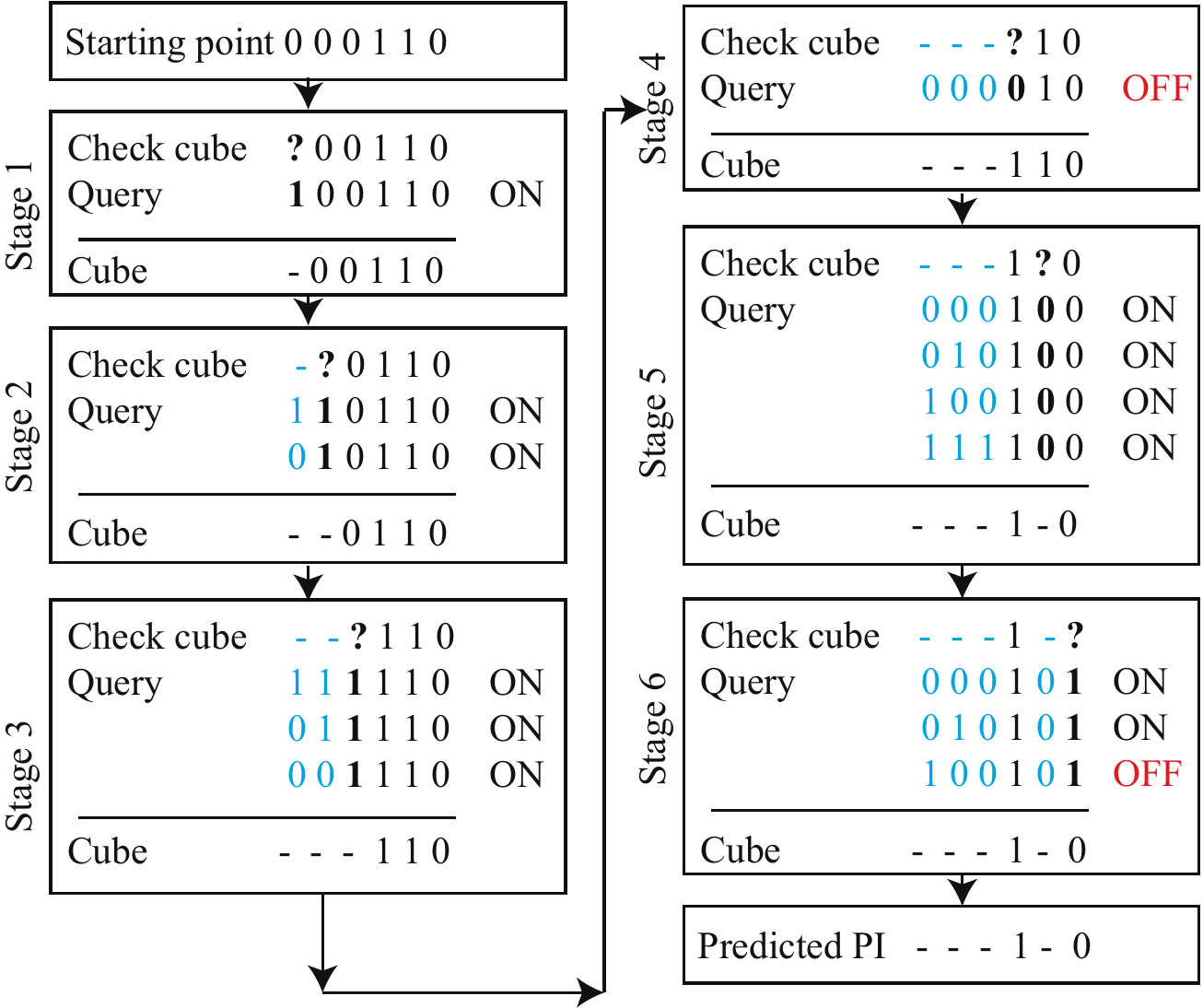} 
    \caption{The PI expansion from
    the initial minterm
    (starting point/seed)
    $m_0=000110$ to $PI_{pred} =\scalebox{1.3}{-}\scalebox{1.3}{-}\scalebox{1.3}{-}1\scalebox{1.3}{-}0$.}
    \label{fig:motivexample}
\end{figure}

We now 
explain
Algorithm~\ref{alg:minterm2PI} with the aid of an example.
Consider the minterm $m_0=000110$, which belongs to 
$f$'s ON-set,
as shown in Figure~\ref{fig:original_netlist_and_pit}. We expand this minterm to a PI in $6$ stages,
as shown in Figure~\ref{fig:motivexample} and described below:

\begin{enumerate}[leftmargin=*] 
    \item Initially, the 
    first
    literal $a_1$
    is flipped to obtain a query $\mathbf{1}00110$. The oracle returns ``$1$'', i.e., the queried minterm belongs to the ON-set. As a result, 
    $a_1$ is replaced with a don't care ``$\scalebox{1.3}{-}$'' and the cube is updated as $\scalebox{1.3}{-}00110$.
    
    \item Then, the 
    second
    literal $a_2$
    is flipped, and the don't care ($a_1$) is replaced by ``$1$'' and ``$0$'' to obtain the next set of queries.
    Recall that we use the linear limitation parameter ($p = 1.1$) to limit the number of queries.\footnote{Based on our tested designs,
    $p=1.1$ is an empirically derived constant.} This implies that we replace $a_2$ to obtain $1.1\times d \approx 2$ queries: $1\mathbf{1}0110$ and $0\mathbf{1}0110$. Since both queried minterms belong to the ON-set,
    $a_2$ is replaced by ``$\scalebox{1.3}{-}$'' to update the cube as $\scalebox{1.3}{-}\scalebox{1.3}{-}0110$.
    
    \item Now, $a_3$
    is flipped, and the two don't cares ($a_1$, $a_2$) are replaced randomly by ``$1/0$'' to obtain $1.1\times d \approx 3$ queries, as shown in Figure~\ref{fig:motivexample}. After observing the result, $a_3$ is replaced by a don't care, and the resultant cube is $\scalebox{1.3}{-}\scalebox{1.3}{-}\scalebox{1.3}{-}110$.
    
    \item Next, $a_4$
    is flipped to ``$0$'', and the don't cares are replaced, as in the earlier stages. The first query
    $000\mathbf{0}10$
    returns ``$0$'', i.e., the queried minterm is OFF-set. Therefore, $a_4$ = $1$, and cube ($\scalebox{1.3}{-}\scalebox{1.3}{-}\scalebox{1.3}{-}110$) is restored back, as shown in Figure~\ref{fig:motivexample}. 
    
    \item Then, $a_5$
    is flipped, and $1.1 \times d \approx 4$ queries are generated by replacing the three don't cares, as shown in Figure~\ref{fig:motivexample}. As all queries belong to 
    the ON-set,
    the cube is updated to $\scalebox{1.3}{-}\scalebox{1.3}{-}\scalebox{1.3}{-}1\scalebox{1.3}{-}0$ by substituting $a_5$ with ``$\scalebox{1.3}{-}$''.
    
    \item In the last step, $a_6$
    is flipped to ``$1$''. The oracle is queried by replacing don't care bits as earlier. However, the third query 
    $10010\mathbf{1}$
    returns ``$0$''. As a result, the step terminates by retaining the cube $\scalebox{1.3}{-}\scalebox{1.3}{-}\scalebox{1.3}{-}1\scalebox{1.3}{-}0$, forming the predicted PI.
\end{enumerate}

The time complexity of Algorithm~\ref{alg:minterm2PI}
is $O(n^2)$.
Thus,
scalability is a concern when $n$ is large.
To reduce the worst-case time complexity, we rely on the following observation: In most practical circuits, not all inputs participate in the computation of a given output. 
Thus, effective inputs\footnote{An effective input can propagate its value to the output.} of the output are usually much less than all inputs.
If these effective inputs can be identified for each output,
then the time taken for 
PI expansion
can be reduced by querying fewer minterms.
Appendix~\ref{sec:proposed_scable_expansion} provides an algorithm and more details on how to scalably expand from an ON-set minterm to a predicted PI.

\subsection{Searching for the Next ON-set Minterm} \label{sec:proposed_single_cone}

After expanding an ON-set minterm to a predicted PI, \proposed~searches for another ON-set minterm to be expanded as the next PI. Rather than randomly searching for the next ON-set minterm in the
complete Boolean 
space ($\mathbb{B}^n$),
\proposed~uses
a satisfiability (SAT)
solver
to reduce the search for the next ON-set minterm. Specifically, the search space is encoded with the following constraints:

\begin{figure}[b!]
    \centering
    \includegraphics[clip, trim=0.8cm 0.6cm 0.6cm 0.8cm, width=0.35\textwidth]{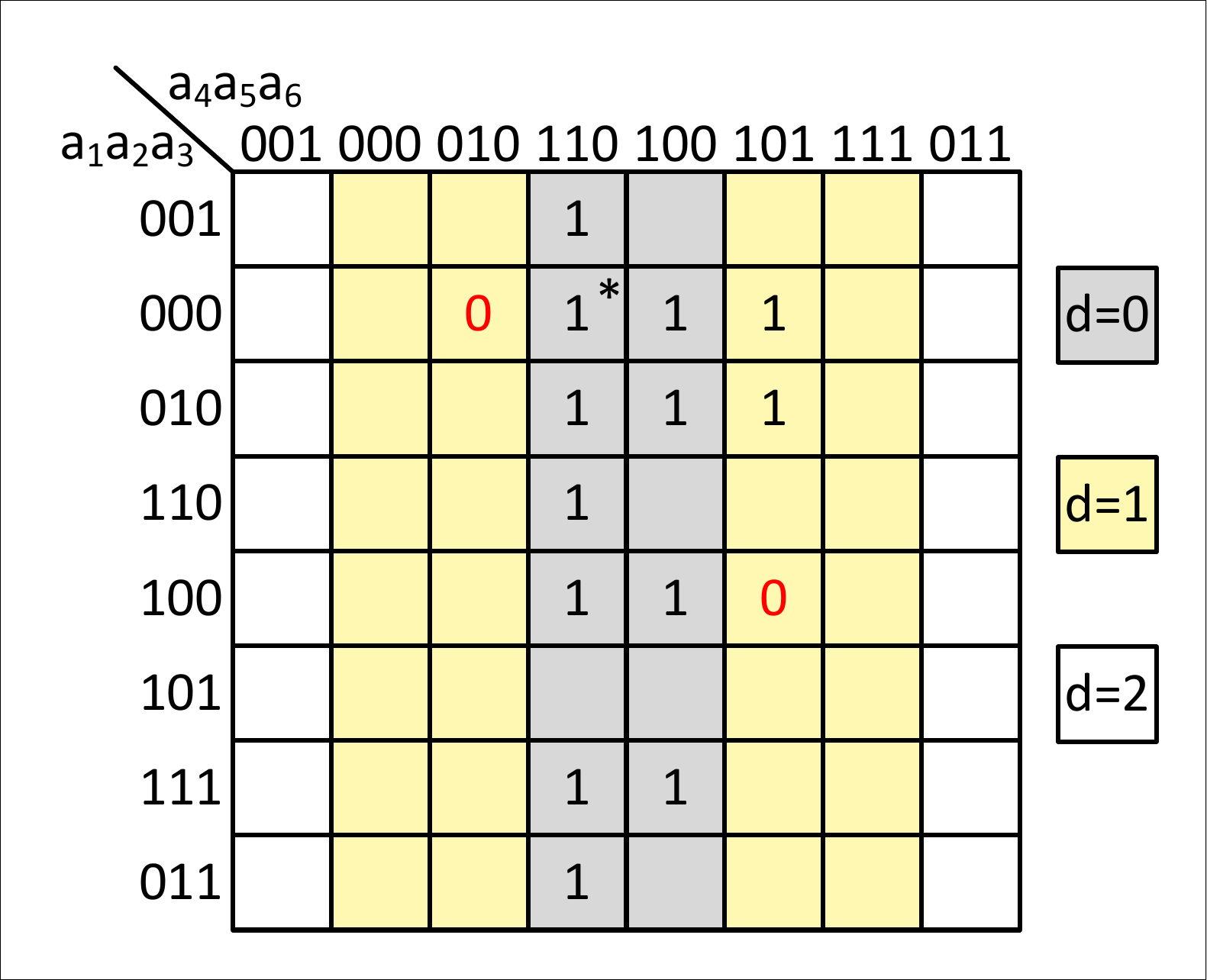} 
    \caption{Search
    space for 
    the
    next ON-set 
    minterm when the current predicted PIT is $\{\scalebox{1.3}{-}\scalebox{1.3}{-}\scalebox{1.3}{-}1\scalebox{1.3}{-}0\}$, as shown in the example of Figure~\ref{fig:motivexample}.
    Gray
    shade denotes
    the space covered by the current predicted PIT, yellow
    shade denotes space with 
    $d=1$
    from PIT, and
    the entry with $*$ denotes the starting point $m_0$.}
    \label{fig:motivnextminterm}
\end{figure}

\begin{itemize}[leftmargin=*]
    \item[(1)] The search 
    space for the next ON-set minterm 
    excludes
    the current predicted PIT. In other words, the next minterm 
    is not within the distance of $d_0 = 0$ of any existing predicted PI. The gray shade in Figure~\ref{fig:motivnextminterm} showcases the excluded space when the current predicted PIT is $\{\scalebox{1.3}{-}\scalebox{1.3}{-}\scalebox{1.3}{-}1\scalebox{1.3}{-}0\}$.
   
    \item[(2)] 
    The search space for the next ON-set minterm only includes the neighboring minterms of the current predicted PIT. More specifically, the search space is bounded within the distance of $d_0$ to the current predicted PIT.\footnote{Distance is the number of pairs of $(0,1)$ and $(1,0)$ between two PIs. The concept of distance is similar to ``Hamming distance,'' except
    distance takes into account PIs containing don't cares, but Hamming distance does not.} The yellow shade in Figure~\ref{fig:motivnextminterm} showcases this constraint when $d_0=1$.
\end{itemize}

For an $n$-input function, if the current predicted PIT has one PI with $x$ don't care bits, the size of the search space for the next ON-set minterm is $(n-x)\times 2^x$ instead of $2^n$. In Figure~\ref{fig:motivnextminterm}, $n=6$ and $x=4$.

We now elaborate on
the encoding process for the next ON-set minterm search according to constraints (1) and (2).

First, we construct a conjunctive normal form (CNF) $CNF_0$ describing the inverse of the predicted PIT.
In other words, the next candidate minterm $m'$ belongs to the OFF-set of the current predicted function $f_{pred}$.
Thus, $CNF_0$ is encoded as
\begin{equation}
    \{ m' \mid f_{pred}(m') = 0 \}\text{.}
    \label{eq:eq1}
\end{equation}

Next, we add 
constraint (2)
to walk through the uncovered space incrementally.
This is done by searching within the distance of $d$ from the PIs in the predicted PIT. The distance
$d$ is incremented from $2$ to $n$ 
(the number of inputs)
to ensure that the search is efficiently directed. Therefore, the candidate minterms belong to the following set
\begin{equation}
    \{m'|\exists m_0\in \mathbb{B}^n, \,
    \text{s.t.}\,
    f_{pred}(m_0) = 1 \text{ and } D(m', m_0)\leq d\}\text{.}
    \label{eq:eq2}
\end{equation}
This CNF encoding is denoted as $CNF_{cstr}^d$.
The constraint CNF for the next ON-set minterm search, $CNF_{search}$, is
\begin{equation}
   CNF_{search} = CNF_0 \wedge CNF_{cstr}^d \text{.}
   \label{eq:eq3}
\end{equation}

These constraints are fed to a SAT solver that returns a candidate minterm $m'$. This minterm ($m'$) is queried using the oracle. If $m'$ belongs to the ON-set, then the search is 
terminated,
and the minterm $m'$ is expanded to
a predicted PI
as described in Algorithm~\ref{alg:minterm2PI}. On the other hand, if $m'$ belongs to the OFF-set, 
the
SAT solver is called again to return another candidate minterm. However, to avoid SAT solver returning the same solution,
$CNF_0$ is updated to exclude the previous result.
Assume 
$m'_{OFF}$ is the previous OFF-set minterm,
then the CNF
constraint in Equation~(\ref{eq:eq1})
is updated as 
\begin{equation}
    CNF_0 = CNF_0 \wedge (\lnot m'_{OFF})\text{.}
\end{equation}

The process of generating candidate minterms is recursive until either finding an ON-set minterm or consecutively visiting $p_{conv}$ OFF-set minterms ($p_{conv} \in \mathbb{N}^+$). $p_{conv}$ is a user-defined convergence parameter for balancing between efficiency and accuracy, as shown in Algorithm~\ref{alg:attack_on_single_cone}. 
Appendix~\ref{sec:appendix_parameters_accuracy} details the analysis on $p_{conv}$.
If $p_{conv}$ OFF-set minterms are consecutively visited, then the constraint $CNF_{cstr}^d$ is updated by incrementing $d$ in Equation~(\ref{eq:eq2}). If $d=n+1$ (the first time that $d>n$), \proposed~terminates.

\begin{algorithm}[t!]
\caption{Prediction on the $w$-th output
}
\label{alg:attack_on_single_cone}

\small

\SetKwFunction{FMain}{\text{predict\_cone}}

\SetKwProg{Fn}{Function}{:}{}

\SetKwProg{IFwithnoend}{if}{ then}{}
\SetKwProg{ELSEwithnoend}{else}{}{}
\SetKwProg{WHILEwithnoend}{while}{ do}{}
\SetKwProg{FORwithnoend}{for}{ do}{}

\SetAlgoLined
\KwIn{
Oracle $\mathcal{O}$, 
output index $w$,
distance parameter $d_0$, 
linear parameter $p$,
{
convergence parameter $p_{conv}$},
time limit $T$
}
\KwOut{Predicted PIT $PIT_{pred}$}

\Fn{ \FMain {$\mathcal{O}$, $w$, $d_0$, $p$, $p_{conv}$, $T$}}{
$PIT_{pred} := \varnothing$ \\
$m_0^{1st} := \text{search\_for\_1st\_ON\_set\_minterm} (\mathcal{O})$\\
$PI_{pred}^{1st} := \text{expand\_minterm\_to\_1stPI} (\mathcal{O}, w, m_0^{1st},)$\\
$PIT_{pred} := PIT_{pred} \bigcup \{PI_{pred}^{1st}\}$\\
\WHILEwithnoend{exe\_time < T}{                     \label{line:another_pi}
  $CNF_0 :=\text{exclude\_PIT}(PIT_{pred})$ \\
  $d := d_0$  \\
  $CNF_{search} := CNF_0 \wedge CNF_{cstr}^d$ \\   \label{line:updated_d}
  $solution := \text{query\_SAT\_solver} (CNF_{search})$ \\
  $counter := 0$  \\
  \WHILEwithnoend{$solution \neq UNSAT$ \& $exe\_time<T$}{
    \IFwithnoend{ $counter \ge p_{conv} $}{
        {\bf break} \hfill {$\rhd$Consecutively visit $p_{conv}$  OFF-set minterms}\\
    } 
    $m' := \text{extract\_minterm}(solution)$ \\
    \IFwithnoend{$\mathcal{O}.\text{query\_oracle} (m_0,w)==1$}{
        $m_0:=m'$ \\
        $PI_{pred} :=~\text{expand\_minterm\_to\_PI} (\mathcal{O}, w, m_0, p)$\\
        $PIT_{pred} := PIT_{pred} \bigcup \{ PI_{pred} \}$ \\ 
        {\bf goto} line \ref{line:another_pi} \\
    }
    \ELSEwithnoend{}{
        $counter := counter+1$ \\
        $m'_{OFF}:=m'$\\
        $CNF_{search} := CNF_{search} \wedge (\lnot m'_{OFF})$ \\
        $solution := \text{query\_SAT\_solver}(CNF_{search})$
    }
  }
  \IFwithnoend{exe\_time < T}{
    $d := d + 1$ \\
    \IFwithnoend{$d==n+1$}{
    {\bf return} $PIT_{pred}$ \\
    }
    \ELSEwithnoend{}{
    {\bf goto} line \ref{line:updated_d}
    }
  }
}
}
{\bf {return}} $PIT_{pred}$ \\
\end{algorithm}

\subsection{Recovering Single Output}
\label{sec:recover_single_cone}
The process of 
searching for
the next
ON-set minterm and the subsequent PI expansion is repeated till 
one of the termination conditions is met.
One termination condition is $d=n+1$, as described at the end of Section~\ref{sec:proposed_single_cone}.
Another termination condition is reaching a user-defined time limit $T$. 
When the time exceeds the
specified time limit,
the attack terminates and returns the predicted PIT generated during attack execution, as described in Algorithm~\ref{alg:attack_on_single_cone}. 
The 
time limit
termination condition is checked after 
each
PI expansion and during the search
for
the next ON-set minterm.

\noindent
{\bf Special Case: Outputs with Constant Logic $\mathbf{0}/\mathbf{1}$.}
The numbers of input and output pins  utilized in an eFPGA design
are
defined by the function being implemented. The remaining 
pins
are constant $0$/$1$. If we can identify these 
constant $0$/$1$ outputs,
then it allows us to allocate more time to predict other non-constant 
outputs.
If the first predicted PI ($PI_{pred}^{1}$) consists of all don't cares (``$\scalebox{1.3}{-}$'') and with no 
literal
(``$0$''/``$1$''),
we regard the 
output's
functionality as constant $1$. On the other hand, if the algorithm cannot find the 
first
ON-set minterm, 
we consider the 
output's
functionality as constant $0$. If the output is determined as
constant $0$/$1$,
then the prediction terminates.

\subsection{Recovering the Entire Circuit} \label{sec:proposed_on_entire}

\begin{algorithm}[t]
\caption{Predicting entire circuit's functionality}
\label{alg:all_cones}

\small

\SetKwFunction{FMain}{predict\_circuit}
\SetKwFunction{AttackOnCone}{\text{predict\_cone}}

\SetKwProg{Fn}{Function}{:}{}

\SetKwProg{FORwithnoend}{for}{ do}{}

\SetAlgoLined

\KwIn{
Oracle $\mathcal{O}$, 
distance parameter $d_0$, 
linear parameter $p$, 
convergence parameter $p_{conv}$, 
time limit $T$
}
\KwOut{Entire predicted circuit $\mathcal{C}_{pred}$}

\Fn{ \FMain {$\mathcal{O}$, $d_0$, $p$, $p_{conv}$, $T$}}{

$output\_size := \text{count\_number\_of\_outputs}(\mathcal{O})$\\
$Cones_{pred} := \varnothing$ \\
\FORwithnoend{$w \in \{1, 2, \cdots, output\_size \}$}{
    $PIT_{pred} := $ \AttackOnCone {$\mathcal{O}$, $w$, $d_0$, $p$, $p_{conv}$, $T$} \\ 
    $Cones^w_{pred} :=~\text{convert\_PIT\_to\_netlist}(PIT_{pred})$\\
}
$\mathcal{C}_{pred} := \text{merge\_cones\_to\_circuit} (Cones_{pred})$ \\
}
{\bf {return}} $\mathcal{C}_{pred}$ \\
\end{algorithm}

Having described how to recover a single output, we now discuss how to recover
the entire circuit.
The prediction on each output
can be made in parallel, as shown in Algorithm~\ref{alg:all_cones}. Once the predictions on all outputs finish, we collect all predicted PITs.
Electronic design automation (EDA)
tools can synthesize the design, such as 
\textit{Synopsys Design Compiler}~\cite{designcompiler}, \textit{Cadence Genus}~\cite{genus}, and \textit{Siemens Precision RTL}~\cite{precisionRTL}, to 
satisfy the desired PPA constraints for the predicted circuit.

\subsection{\proposed~in IC and FPGA Design Flows}
\label{sec:design_flow_after_prediction}

\begin{figure}[t]
    \centering
    \includegraphics[clip, 
    trim=0.65cm 0.61cm 0.56cm 0.65cm, width=0.46\textwidth]{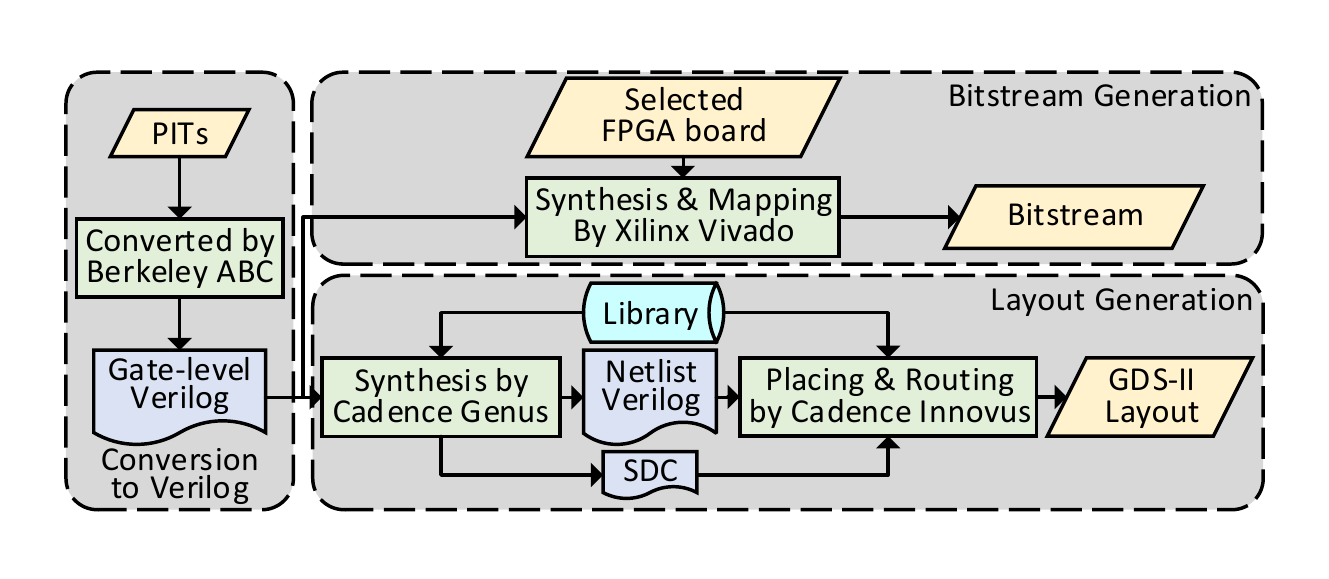} 
    \caption{The IC design flow with \proposed-predicted PITs.}
    \label{fig:post_design_flow}
\end{figure}

\noindent
Since the predicted logic format is PIT, it raises the question of \textit{how to utilize the predicted design in the IC design flow}. We address this by discussing the steps the attacker can take after using \proposed.
After recovering an eFPGA-redacted design, the attacker can flexibly choose to either upload the bitstream with predicted design information on the FPGA or replace the redacted components with ASIC netlist and generate its layout, as shown in Figure~\ref{fig:post_design_flow}. 
To generate the bitstream of the FPGA or the layout of the ASIC, the attacker could utilize academic tools (e.g., \textit{Berkeley ABC}~\cite{ABC}) and industry-standard commercial tools (e.g., 
\textit{Xilinx Vivado}~\cite{DisableBitstreamXilinx}, \textit{Synopsys Design Compiler}~\cite{designcompiler}, \textit{Cadence Genus}~\cite{genus}, and \textit{Cadence Innovus}~\cite{CadenceInnovus}.
Next, we describe in detail the bitstream and layout generation.

\noindent
\textbf{Bitstream Generation.}
\proposed~predicts PITs of the target design with input-output query access. 
After collecting the predicted PITs of all outputs, we convert them into a structural netlist by 
converting
the logic of each output’s PIT 
into a Verilog file using
\textit{Berkeley ABC}~\cite{ABC}.
This gate-level design
is then passed on to 
\textit{Xilinx Vivado}~\cite{DisableBitstreamXilinx}
for synthesis, implementation, and bitstream generation.
Thus, 
the attacker can upload this bitstream of the predicted design on the FPGA.

\noindent
\textbf{Layout Generation.}
Similar to the process of generating bitstream, the attacker can convert the predicted PITs to 
the
Verilog design,
as shown in Figure~\ref{fig:post_design_flow}.
Then, we use \textit{Cadence Genus}~\cite{genus}
to generate the synthesized
netlist Verilog design with the selected library,
as shown in Figure~\ref{fig:post_design_flow}. 
In this synthesis process, \textit{Cadence Genus} also 
merges/optimizes the logic and
generates an
industry-standard SDC file containing design constraints and timing assignments. 
Further, to generate the layout, we provide the netlist Verilog file, SDC file, and library as inputs to 
\textit{Cadence Innovus}~\cite{CadenceInnovus},
the physical implementation tool. 
\textit{Cadence Innovus} optimizes the placing and routing processes to generate the optimal layout.
The timing, power, and size characteristics of the \proposed-recovered design will vary and depend on what synthesis tools and constraints are used by the attacker and the defender. eFPGAs are purported to protect functionality~\cite{bandari2021iccad,mohan2021hardware}, which is the challenge we target in this work.

\section{Results}
\label{sec:simu}
In this section, we run \proposed~on select practical circuits and discuss the efficiency and effectiveness of \proposed~compared to the state-of-the-art technique, Chen \textit{et al.}~\cite{chen2020circuit}. Later, we verify the validity of the underlying assumptions of our attack and corroborate these assumptions with our results. Finally, we further analyze \proposed's performance on a few selected circuits and present them as case studies for more in-depth understanding.

\subsection{Simulation Setup}   \label{sec:setup}
\proposed~targets practical circuits considering only black-box access to the combinational part of a design.
Specifically, \proposed~attacks hardware designs redacted by embedded field-programmable gate arrays (eFPGAs).
To reduce the impact of inherent variation among multiple vendors due to factors such as 
field-programmable gate array (FPGA)
board size and clock speed, we evaluate \proposed~by testing it against circuits implemented on the OpenFPGA framework~\cite{openfpga}.

\noindent {\bf Environment:} We run the attack simulations on a 32-core Intel Xeon processor at 2.6 GHz with 512 GB RAM.
We use {\it Verilator}~\cite{verilator} for generating executable binary oracles of tested circuits,
{\it Synopsys Design Compiler}~\cite{designcompiler} for synthesis, and the {{\it Berkeley ABC}} tool~\cite{ABC} for converting to Verilog.

\noindent{\bf Tested Circuits:}
We select a wide range of test circuits at different scales, including seven
circuits from ISCAS'85 and ITC'99 benchmark suites, 
two processor circuits (Stanford MIPS and IBEX), 
and the Common Evaluation Platform (CEP) GPS circuit~\cite{iscas85,itc99,hennessy1981mips,woods2014ibex,darpa_cep}.
Further, we evaluate \proposed's performance on
circuits from the Cybersecurity Awareness Worldwide (CSAW) 2021 competition~\cite{csaw2021}.\footnote{CSAW is the largest student-run cybersecurity event in the world~\cite{csaw2021}.}

\begin{table*}[bht!]
    \centering
    \caption{A performance comparison between
    Chen \textit{et al.}~\cite{chen2020circuit} and \proposed~on multiple 
    test
    circuits using different metrics.
    Some cells are marked as ``{\it erroneous case},'' which implies the failure of Chen {\it et al.}~\cite{chen2020circuit} to run on the tested circuit.}
     \label{tab:scale_and_results_benchmark_and_processor}
    \resizebox{0.95\textwidth}{!}{
    \begin{tabular}{|c|c|c|c|c|c|c|c|c|c|}
    \hline
    \multicolumn{2}{|c|}{
    \multirow{3}{*}{\bf Circuit}} & \multirow{3}{*}{\bf \# inputs} & \multirow{3}{*}{\bf \# outputs} & \multirow{3}{*}{\bf \# gates} 
    & \multicolumn{4}{c|}{\bf Success rate} & {\bf Efficiency} \\ \cline{6-10}
    \multicolumn{2}{|c|}{} &&&& \multicolumn{2}{c|}{\bf Formal (\%)} & \multicolumn{2}{c|}{\bf Simulation (\%)} & {\bf Attack time (hour)} \\ \cline{6-10}
    \multicolumn{2}{|c|}{} &&& & {\bf Chen \textit{et al.}~\cite{chen2020circuit}} & {\bf \proposed} & {\bf Chen \textit{et al.}~\cite{chen2020circuit}} & {\bf \proposed} & {\bf \proposed} \\ \hline\hline
    
    \multirow{7}{*}{\rotatebox{90}{\bf Benchmark } }& c432 &$36$ &$7$ & $160$ 
    & $0$ & $28.57$
    & $82.51$ & $99.86$ 
    & $0.09$
    \\ \cline{2-10}
    & c880 &$60$ &$26$ & $383$
    & $0$ & $53.85$
    & $82.01$ & $96.12$ 
    & $1.06$
    \\ \cline{2-10}
    & c1355 &$41$ &$32$ & $546$ 
    & $0$ & $0$
    & $61.02$ & $50.77$ 
    & $1.01$
    \\ \cline{2-10}
    & c1908 &$33$ &$25$ & $880$ 
    & $0$ & $0$
    & $63.73$ & $81.82$ 
    & $1.00$
    \\ \cline{2-10}
    & c7552 &$207$ &$107$ & $3,512$ 
    & $0$ & $53.27$
    & $72.05$ & $86.66$ 
    & $1.25$
    \\ \cline{2-10}
    & b14 &$277$ &$299$ & $9,821$ 
    & {\it erroneous case}
                & $20.40$
    & {\it erroneous case} & $92.46$    
    & $1.39$
    \\ \cline{2-10}
    & b20 &$522$ &$512$ & $6,787$ 
    & $0$ & $11.91$
    & $64.26$ & $84.10$   
    & $9.92$
    \\ \hline \hline
    \multirow{3}{*}{\rotatebox{90}{\bf Others }} &MIPS 
    & 
    {$466$} 
    & 
    {$330$}
    & $3,902$ 
    & $0$ & $63.39$
    & $81.97$ & $95.49$   
    & $1.84$
    \\ \cline{2-10}
    & IBEX & $1,386$ & $1,385$ & $18,087$ 
    & $0$ & $16.46$
    & $56.68$ & $90.96$ 
    & $72.85$
    \\ \cline{2-10}
    & GPS & $9,707$ & $9,731$ & $213,125$ 
    & {\it erroneous case}  & $25.36$ 
    & {\it erroneous case} & $68.89$
    & $44.44$ 
    \\ \hline
    \end{tabular}
    }
\end{table*}

\subsection{Evaluation Metrics}

\label{sec:metric}
To assess the performance of \proposed, we employ the following two 
metrics:
(i) equivalence checking and
(ii) an accuracy metric 
quantifying the functional similarity
between the predicted and original circuits.

\noindent\textbf{Equivalence Checking:}
This metric considers the ratio of functionally equivalent outputs to all outputs.
Commercial tools such as {\it Synopsys Formality}~\cite{SynopsysFormality} and {\it Cadence Conformal}~\cite{CadenceConformal}
are used to perform the equivalence check.
The equivalence checking tool identifies whether a predicted output is equivalent to the original circuit. 
Naturally, the tool does not quantify the error rate
or functional similarity. 
Thus, this metric may be misleading as a design may have a very low error rate (<0.01\%) but may have a low equivalence check score. Thus, we use another metric to quantify the error rate/accuracy and provide an in-depth evaluation.

\noindent{\bf Simulation-based Accuracy:}
Due to the drawback of the equivalence checking-based metric, we need to formalize an output accuracy-based metric to ensure a fair comparison with previous work and correctly measure attack effectiveness.
To this end, we employ a simulation-based method to quantify the functional similarity between the original and predicted designs.
For every
$i^{th}$ output ($i\in\{1,2,\cdots, m\}$, where there are $m$ outputs),
we test the original and predicted circuits with a set of 
randomly chosen minterms and record the number of minterms with matched output responses from both circuits.
Thus, we calculate the accuracy
of
the $i^{th}$ output as
\begin{equation*}
AC_i = \frac{\text{\# minterms with matched responses}}{\text{\# tested minterms}} \times 100\%\text{.}
\end{equation*}
We repeat this process on all outputs and take the average value $AC = \sum\limits_{i=1}^{m} AC_i \! \big/ \! \! m$ 
as the accuracy of the entire design.

\subsection{Performance of \proposed}   
\label{sec:attacks_on_practical_circuits}

We use the following parameter values to run \proposed:
distance parameter $d_0=2$, 
linear parameter $p=1.1$,
constant limitation parameter $p_0=8$,
and convergence parameter $p_{conv}=50$.
We obtain these values empirically.

Table~\ref{tab:scale_and_results_benchmark_and_processor} shows that \proposed~has a higher accuracy compared to the state-of-the-art technique, Chen \textit{et al.}~\cite{chen2020circuit}, on all tested circuits except for c1355. 
The average accuracy (simulation-based) of Chen \textit{et al.}~\cite{chen2020circuit} and \proposed~is $70.5$\% (excluding b14 and GPS) and $84.7$\%, respectively. Thus, \proposed~has an improvement of $14.2$\% compared to Chen \textit{et al.}~\cite{chen2020circuit}. 
\proposed~achieves an accuracy of $70\%$ while attacking GPS, with an attack time of $44.4$ hours. 
Meanwhile, Chen \textit{et al.}~\cite{chen2020circuit} fails to run on b14 and GPS, while it runs for more than $5$ days on IBEX and achieves an accuracy of only $56.68$\%.
We believe these anomalies are due to mismatches between the circuits and the tool's presumptions.
Chen \textit{et al.}~\cite{chen2020circuit} relies on similarities in port names and the presence of few selected linear operators, whereas \proposed~makes no such presumptions.
Therefore, \proposed~achieves better 
prediction performance compared to Chen \textit{et al.}~\cite{chen2020circuit} and recovers more designs.

The attack time for \proposed~is the sum of two parts: (i) the maximum execution time while predicting single outputs and (ii) the time taken to merge all predicted single output cones into the entire circuit. For an attacker unconstrained by a lack of computational resources, it may be possible for them to run all single output predictions in parallel to speed up (i). However, considering computational limits, we run our attack on a maximum of 50 outputs in parallel.

In addition to the tested circuits in Table~\ref{tab:scale_and_results_benchmark_and_processor}, we also run Chen \textit{et al.}~\cite{chen2020circuit} and \proposed~on the circuits featured in the CSAW 2021 competition~\cite{csaw2021}.
The logic locking event included circuits redacted with eFPGA based on the OpenFPGA framework. 
For details, please check Appendix~\ref{sec:appendix_csaw}.

\subsection{Survey on Distance Between PIs}

\label{sec:distribution_hd}

\begin{figure}[tb!]
    \centering
    \includegraphics[clip, 
    trim=0.65cm 0.8cm 0.75cm 0.65cm,
    width=0.35\textwidth]{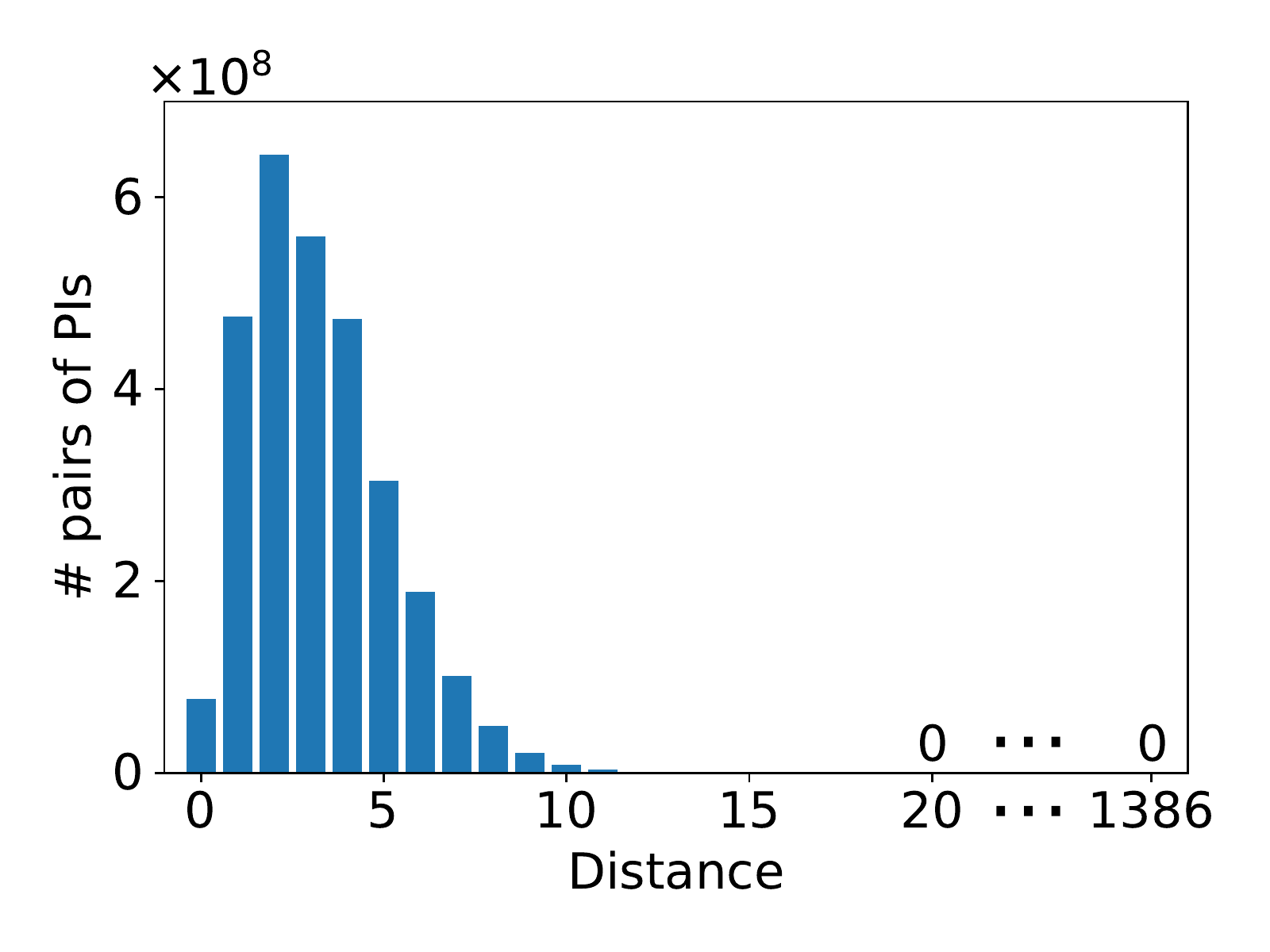}
    \caption{
    The distribution of the distance between two PIs of the extracted PIT for the case study of  IBEX~\cite{woods2014ibex}.
    }
    \label{fig:distribution_of_hd}
\end{figure}

As mentioned in Section~\ref{sec:attacks_on_practical_circuits}, we choose the distance parameter $d_0=2$ in \proposed. 
In this subsection, we explain the reason behind this decision.
The choice of $d_0=2$
is due to the distances between prime implicants (PIs) 
on most practical hardware intellectual properties (IPs).
To understand these distance properties on practical circuits, we 
conduct a study on IBEX~\cite{woods2014ibex} 
and 
investigate
the prime implicant table (PIT) of each output using the {\textit{Berkeley ABC}} tool.
First, for each PIT, we calculate the distance between any two PIs within the PIT.
Then, we collect the distance distribution among all extracted outputs PITs, as shown in Figure~\ref{fig:distribution_of_hd}.
In each extracted PIT, all pairs of PIs are with a distance of $\le 19$. 
This result gives a preliminary idea of the distance distribution for each pair of PIs. 
Further, to show how close each PI is to the rest of the PIT, we investigate the minimum distance of each PI to all other PIs in the same PIT, as shown in Table~\ref{tab:distribution_of_min_distance}.
Table~\ref{tab:distribution_of_min_distance} shows
that the minimum distance between each PI to other PIs in the same PIT is 
either $0$, $1$, or $2$. In other words, in the same PIT, for an arbitrary PI $PI_i$, there always exists another PI $PI_j$ in the same PIT, such that $D(PI_i, PI_j)\le 2$. Thus, to make the process of searching for the next ON-set minterm efficient and effective,
we choose $d_0=2$. 
The findings of this analysis are generalizable to most practical circuits as 
the findings
are an outcome
of 
electronic design automation (EDA)
algorithms clustering minterms together during the design encoding stage.
Thus, the choice of $d_0=2$ is valid for most practical hardware IPs.

\begin{table}[t!]
    \centering
    \caption{A case study of IBEX~\cite{woods2014ibex}: the distribution of the minimum distance for PIs.}
    \resizebox{0.48\textwidth}{!}{
    \begin{tabular}{|c||c|c|c|c|cc|c|}
    \hline
    {\bf Distance}
    &  $0$ &  $1$  &   $2$ &   $3$  & \multicolumn{2}{c|}{$\cdots$} &  $1,386$ \\ \hline 
    {\bf \# PIs} & $7.8\times 10^{5}$  &  $1.4\times 10^{3}$  & $4$ & $0$ &\multicolumn{2}{c|}{$\cdots$} &$0$ \\ \hline
    \end{tabular}
    }
    \label{tab:distribution_of_min_distance}
\end{table}

\begin{figure}[bth!]
    \centering
    \includegraphics[clip, 
    trim=0.6cm 0.7cm -0.13cm 0.7cm, width=0.4\textwidth]{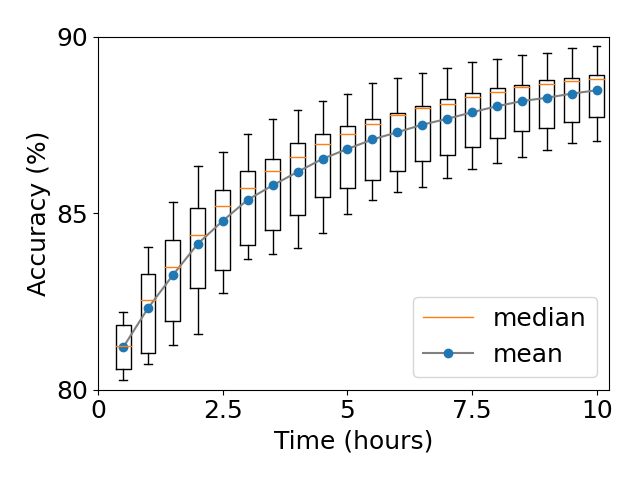}
    \caption{The performance trade-off between accuracy and the attack time of \proposed~on  IBEX~\cite{woods2014ibex}.}
    \label{fig:boxplot_tradeoff_time_vs_AC_IBEX}
\end{figure}

\subsection{Performance Trade-off}
\label{sec:trade-offs}

To observe the trade-off between the time limit and the accuracy, we repeat \proposed~on IBEX~\cite{woods2014ibex} $10$ times. We choose IBEX as the design to be tested as  IBEX has $1,386$ inputs and $1,385$ outputs and can be considered a large-scale design. 
Further, we calculate the accuracy with different time limits. The time limit estimates the total attack time. Note that \proposed~terminates PIT predictions within the time limit; however, the actual attack time may exceed the time limit due to 
the last PI expansion still taking place and
the PIT merging 
processes for individual cones.

Figure~\ref{fig:boxplot_tradeoff_time_vs_AC_IBEX} shows the box plot considering $10$ sets of attacks on IBEX when the time limit 
($T$)
ranges from $0.5$ hours to $10$ 
hours with $\Delta T=0.5$ hours as the time step.
When 
$T=0.5$
hours, the average accuracy is $81.2\%$; when 
$T=10$
hours, the average accuracy is $88.5\%$. Among the repeated $10$ attacks on IBEX, the standard 
deviations of accuracy with different time limits are
always within $1.5\%$.
Thus, the accuracy results are stable since the random processes in \proposed~do not result in significant accuracy deviations among repeated attacks.
Further, 
by observing 
the tendency of the accuracy vs. time curve,
the attacker can fit the collected data in a mathematical model, such as a logarithmic model, and use it to estimate the accuracy for a desired time limit $T$, and vice versa.

\subsection{\proposed~on a Real-World Application}
\label{sec:image_functeller}

\begin{figure}[b!]
    \centering
     \begin{subfigure}[b]{0.18\textwidth}
         \centering
         \includegraphics[clip, trim=4cm 2cm 5cm 1cm,
         width=1\textwidth]{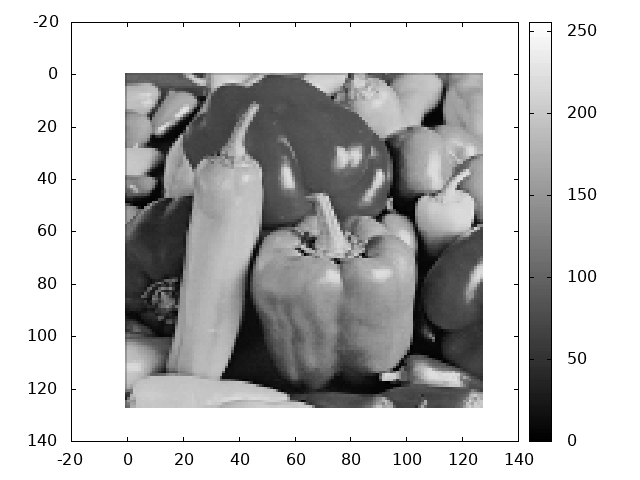} 
         \caption{}
     \end{subfigure}
    \begin{subfigure}[b]{0.18\textwidth}
         \centering
         \includegraphics[clip, trim=4cm 2cm 5cm 1cm,
         width=1\textwidth]{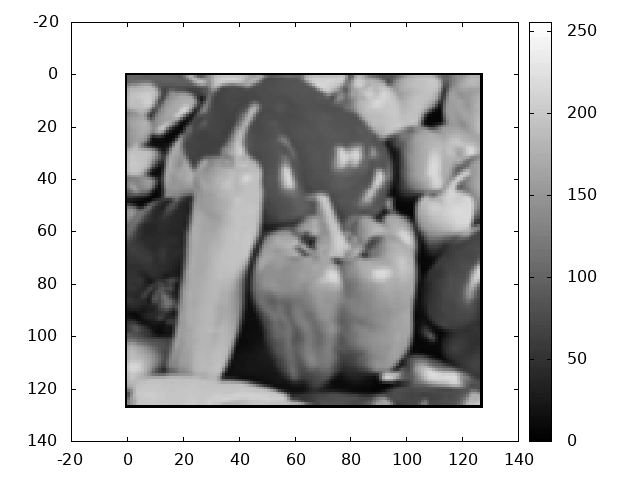} 
         \caption{}
     \end{subfigure}
     \begin{subfigure}[b]{0.18\textwidth}
         \centering
         \includegraphics[clip, trim=4cm 2cm 5cm 1cm,
         width=1\textwidth]{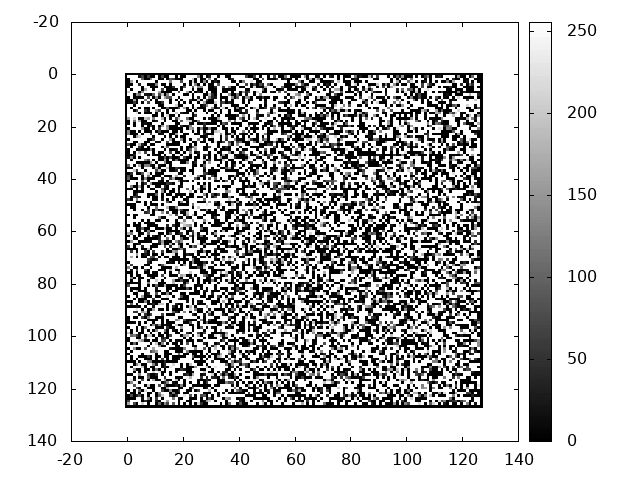} 
         \caption{}
     \end{subfigure}
     \begin{subfigure}[b]{0.18\textwidth}
         \centering
         \includegraphics[clip, trim=4cm 2cm 5cm 1cm,
         width=1\textwidth]{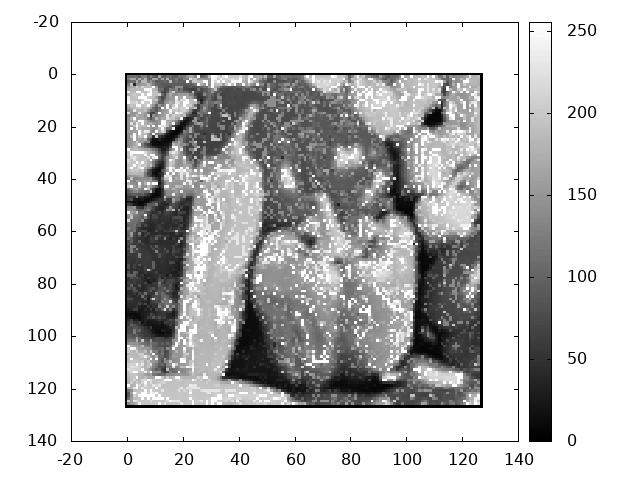} 
         \caption{}
     \end{subfigure}  
    \caption{(a) Input image, (b) original output image, (c) output image when all the $16$-bit adders are redacted by eFPGA, and (d) output image after the recovery of \proposed.}
    \label{fig:Peppers_image}
\end{figure}

\noindent 
Recall that we evaluate the performance of \proposed~by calculating the accuracy metrics discussed in Section~\ref{sec:metric}. 
We now show that the design recovered by \proposed~can meaningfully work by evaluating the performance of \proposed~at the application level.

We consider a hardware-software co-design application
for image processing~\cite{FPGAImageProcessingPlatform}. The core of this application is a hardware design in Verilog, {\tt image\_processing.v}. This application reads an image and generates an output image based on Gaussian blur, an image processing function.
There are a total of eighteen $16$-bit adders in the image processing circuit. We assume that this circuit is protected by redacting all the $16$-bit adders to an eFPGA. We then use \proposed~to recover the circuit by predicting and replacing the redacted modules.  Thus, the recovered image processing circuit contains eighteen \proposed-recovered $16$-bit adders.
In our evaluation, we compare the output images of:
(i) the golden image processing circuit with programmed eFPGA, (ii) the protected image processing unit with un-programmed eFPGA, and (iii) the recovered image processing circuit with \proposed~prediction.

We use a $128\times128$-pixel image of Peppers for our evaluation, as shown in Figure~\ref{fig:Peppers_image}(a). We perform Gaussian blur image processing on this input. 
Figure~\ref{fig:Peppers_image}(b) shows the output image of the original hardware, which is the golden reference.
Figure~\ref{fig:Peppers_image}(c)
shows the output of the image processing circuit with un-programmed eFPGA. Recall that all eighteen $16$-bit adders are redacted to an eFPGA. To mimic the behavior of the un-programmed eFPGA, we replaced all $16$-bit adders with random number generators. 
The result in Figure~\ref{fig:Peppers_image}(c) shows that 
the
image content is not recognizable. 
Finally, Figure~\ref{fig:Peppers_image}(d) shows the output image of the \proposed-recovered image processing circuit. A comparison of Figure~\ref{fig:Peppers_image}(b) and Figure~\ref{fig:Peppers_image}(d) images shows that the output image of recovered hardware is similar to that of the golden one.

\section{Related Works} 
\label{sec:related}

Application-specific integrated circuits (ASICs) and field-programmable gate arrays (FPGAs) allow for a variety of intellectual property (IP) protection mechanisms.
Since there has been a significant interest in developing IP protection mechanisms, it is essential to determine the impact of \proposed~in the field. To this end, we present related works in IP protection to understand \proposed's relation to the current literature.

\subsection{IP Protections and Attacks on ASIC} \label{sec:previous_countermeausures}

Preventing piracy of ASIC designs led to the genesis of initial IP protection countermeasures. Concurrent efforts by the research community led to the development of multiple defensive countermeasures, where most countermeasures protected against threats emerging from untrusted entities with minor 
differences in their threat models and use-case scenarios.
Notably, almost all developed defenses were also successfully broken by innovative attacks. 

\noindent {\bf Logic Locking} 
inserts additional circuitry along with additional key inputs. 
The functionality is retrieved only with the 
correct
key~\cite{epic, AntiSAT2}.
Thus, logic locking
can defend against threats such as reverse engineering and IP piracy.
Logic locking has gained significant 
interest
as an
IP piracy 
countermeasure.
For instance,
the
Siemens
Security TrustChain platform integrates logic locking as a countermeasure~\cite{TrustChain}.
However,
powerful 
input-output (I/O) query-based and structural attacks~\cite{pramod2015sat,MengliProvablySecureCamo,pramod2019funct}
break
many logic locking techniques.

\noindent {\bf  Camouflaging}~\cite{camoStdCell} 
adds additional dummy layout structures to hide the IP layout details. Notably, Rambus utilized  camouflaging and developed
{its}
{proprietary camouflaging technology}~\cite{RambusICCamou}.
However, a design protected by camouflaging can be mapped to a logic-locked design~\cite{yasin2015transforming}. 
Thus, pirating the hardware IP protected by camouflaging is equivalent to breaking the corresponding logic-locked design~\cite{el2019sat}.

\noindent {\bf Split Manufacturing} is 
an
IP protection technique applied to IP layout.
The layout layers are split into the back-end-of-line (BEOL) and front-end-of-line (FEOL) layers.
Then, the FEOL layers are sent to an untrusted foundry, and a trusted foundry fabricates the BEOL layers.
Without the BEOL connections, the attacker (in the FEOL foundry) cannot directly pirate the design.
The Intelligence Advanced Research Projects Activity (IARPA) proposed split manufacturing to protect chip fabrication~\cite{IARPA_TIC}.
However, there have been successful algorithmic attacks against split manufacturing, such as proximity attacks on 
split-manufactured
designs~\cite{wang2016cat, li2019attacking}.

Multiple proposed attacks defeat IP protection 
techniques on ASICs.
As stated in Table~\ref{tab:general_relationship_threats_countermeasures}, 
algorithmic attacks can circumvent the previously mentioned countermeasures (logic locking, camouflaging, and split manufacturing) on ASICs.
However, algorithmic attacks cannot successfully adapt to eFPGAs. 
I/O attacks, such as
Satisfiability (SAT)
attack~\cite{pramod2015sat} and its 
variants, cannot 
scale to attack eFPGAs. This incompatibility is due to the exponentially increased complexity 
of attacking large 
lookup table (LUT) based
structures
for SAT solvers. Structural 
attacks~\cite{MengliProvablySecureCamo, pramod2019funct}
are not applicable to eFPGAs due to the lack of meaningful structural 
traces since each LUT's function and routing information are not readily available.
Thus collectively, algorithmic attacks, one of the most popular forms of IP theft attacks, fail to break eFPGAs.

\subsection{IP Protection Techniques on FPGA} 

\noindent {\bf Bitstream Protection.}
Previous works have considered extracting 
the design from the bitstream~\cite{bitstream2018,2012bitstream}.
The bitstream of an FPGA is encoded with the design functionality. Thus, bitstream extraction would enable reverse engineering of the 
{design} 
implemented on the eFPGA. Thus, it is 
important
to protect the eFPGA bitstream by developing countermeasures. Bitstream extraction attacks can be circumvented by disabling read-back capabilities, ensuring strong encryption on the bitstream, and storing the bitstream in a tamper-proof memory~\cite{DisableBitstreamXilinx}. Furthermore, other bitstream protections from a hardware perspective~\cite{yang2004scan, szefer2019survey, koeune2005tutorial,SummaryFPGAProtect,AgainstDPA} defend against side-channel attacks. The increased interest in developing bitstream extraction countermeasures calls for newer attack methods to recover the design redacted by eFPGA.

\subsection{Alternative Attacks on eFPGA}  \label{sec:alternative_attacks}
Chhotaray \textit{et al.}~\cite{chhotaray2021hardening} stated that, for a protected design, the main objective should be function recovery rather than recovering the correct key when facing a strong attack model~\cite{chhotaray2021hardening}, such as the threat model in this paper.
Functional recovery considers the ratio of input patterns that result in the correct output and total input patterns. The concept of function recovery is useful in quantifying attacks for eFPGA redacted hardware. 
This paper uses this concept to quantify the performance of 
\proposed, as shown in Section~\ref{sec:metric}.

Recently,
attacks
following this principle have been proposed. Chowdhury~\textit{et al.}~\cite{chowdhury2022predictive} use a predictive model which aims to map the eFPGA redacted design to a previously known circuit.
This approach requires obtaining data from a large pool of circuits~\cite{chowdhury2022predictive}; however, creating such a dataset is challenging, especially for proprietary designs, such as industrial processors, as they are not open-source.
\proposed~does not need such a dataset and, thus, can be applied to proprietary designs.
Recent research also investigates the percentage of redacted fabric in the design. Ulabideen~\textit{et al.}~\cite{ulabideen2022security} found that an obfuscation rate of 80\% on SHA-256 prevents most template-based attacks. Further, they could synthesize the stated design to match current state-of-the-art constraints. \proposed, due to its heuristic approach, remains effective as the attack is independent of the design's obfuscation rate and is not dependent on a dataset of previous designs, which is a drawback of predictive and template-matching approaches.
\section{Discussion}
\label{sec:discussion}
Even if \proposed~broadly breaks most practical hardware designs efficiently and effectively, some limitations exist.

\subsection{Limitations of \proposed} \label{sec:limitation_proposed_attack}

Table~\ref{tab:scale_and_results_benchmark_and_processor} shows that the \proposed~predicts circuit functionality with an average accuracy of
$85$\%.
However, the accuracy drops to close to $50\%$
on c1355 and Advanced Encryption Standard (AES) circuits. Analyzing these corner cases helps develop countermeasures against \proposed.

c1355 is a 32-bit single-error-correcting  design~\cite{iscas85}. This implies 
that,
for any two adjacent minterms 
keeping a distance of $1$,
one of the minterms belongs to the ON-set, and the other belongs to the OFF-set~\cite{hamming1950error}. 
In other words, on each output of c1355, each ON-set minterm is a prime implicant (PI),
and there are  
$2^{32-1}$ PIs in 
each prime implicant table (PIT).
The PIT of each c1355's output contains the exponential number of PIs to the input size, which is not scalable.
As a result, 
\proposed~predicts c1355 with low accuracy,
even though c1355 is small-scale with $41$ inputs, $32$ outputs, and $546$ gates.

AES, a popular and well-researched cryptographic core, is also seemingly secure against \proposed.
To verify this
hypothesis, we choose to test \proposed~on an AES design
for $10$ rounds with an unknown key.
We run
\proposed~on the
AES circuit
with the 
time limit set at $24$ hours per output. 
As a result, \proposed~achieves an accuracy of $49.99\%$ on AES.
When the key is unknown, the AES design is a pseudo-random function.
However, we demonstrate \proposed~performance on most practical circuits, such as processor IPs, which constitute a major share of the hardware IP market, according to a recent semiconductor market analysis~\cite{HardwareIPMarket}.

\subsection{Potential Countermeasures}
\label{sec:potential_countermeasures}

The threat model of \proposed~assumes that the attacker can obtain the unauthorized scan-chain access by performing attacks~\cite{agrawal2008scan,banik2013improved,banik2014cryptanalysis,cui2017current,DaRolt2011scanattackandcountermeaures,ali2015novel,alrahis2019scansat,limaye2020dynunlock}, as described in Appendix~\ref{sec:appendix_scan_chain}.
\textit{ Dishonest oracle (DisORC)}~\cite{limaye2020thwarting} protects the black-box design (oracle) and remains unbroken against various attacks. 
If the attacker tries to enable the scan access, the functionality of the design is corrupted, and hence, the outputs are incorrect.
However,
DisORC may not sufficiently protect a circuit from recovery by \proposed~because DisORC’s effectiveness is limited.
The Hamming distance (HD) between the correct output and the output on applying a random key introduced by DisORC is insufficient~\cite{rajendran2014regaining, limaye2020thwarting}, so the attacker may still use \proposed~to recover functionality, as described in Appendix~\ref{sec:appendix_scan_chain}. Therefore, a potential countermeasure to defeat \proposed~should consider both strong security of scan-chain access and sufficient HD.
Further, a defender can harden the recovery of the redacted designs against \proposed~by carefully selecting and redacting the logic module whose PI distribution profile is similar to that of the AES or c1355. In other words, if the redacted logic PIT has a large number of PIs and literals (e.g., exponential to the input size), the redacted logic is secure against \proposed, as discussed in Section~\ref{sec:limitation_proposed_attack}. Yet, this may also exponentially increase the hardware size.

\section{Conclusion and Ramifications} \label{sec:conclusion}

This paper proposes
a heuristic approach, \proposed,
to extract the approximate functionality of the design 
implemented on an embedded field-programmable gate array (eFPGA). 
\proposed~exploits the attribute of most practical hardware Boolean functions, where the prime implicants (PIs) are close to each other. \proposed~can effectively recover most practical designs, including benchmark circuits and processors.
Significantly, \proposed~achieves $>90\%$ accuracy on the IBEX processor.
However, \proposed~cannot recover pseudo-random functions, such as Advanced Encryption Standard (AES).
In this paper, we have focused our discussion solely on the eFPGA platform, as our goal is to
challenge the security of eFPGA-based hardware redaction; we plan to extend \proposed~to other hardware platforms such as ASICs, FPGAs, and cloud FPGAs.

\section*{Acknowledgement}

We thank Cybersecurity Awareness Worldwide (CSAW) 2021 logic locking competition organizers for providing the competition circuits and platform~\cite{csaw2021}.
We thank the members of the TAMU SETH lab for their help in improving the paper. 
Moreover, we thank the reviewers and Shepherd for providing valuable comments during the reviewing process.
The work was supported in part by 
the National Science Foundation
(NSF CNS--1822848 and \# 2016650) and
the Defense Advanced Research Projects Agency (DARPA) grants (HR0011-20-9-0043 and M2102069) from the Automatic Implementation of Secure Silicon (AISS) program~\cite{darpa_aiss} and the Structured Array Hardware for Automatically Realized Applications (SAHARA) program~\cite{sahara}.
Any opinions, findings, conclusions, or recommendations expressed herein are those of the authors, and do not necessarily reflect those of the US Government.

{\footnotesize
\bibliographystyle{plain_auth_limited} 
\bibliography{Bibfile}
}

\appendix
\section*{Appendix} \label{apd:appendix}

\section{Scan-Chain Access for FPGA Designs}
\label{sec:appendix_scan_chain}

\begin{figure}[b]
    \centering
    \includegraphics[clip, trim= 0.8cm 0.7cm 0.9cm 0.9cm, width=0.43\textwidth]{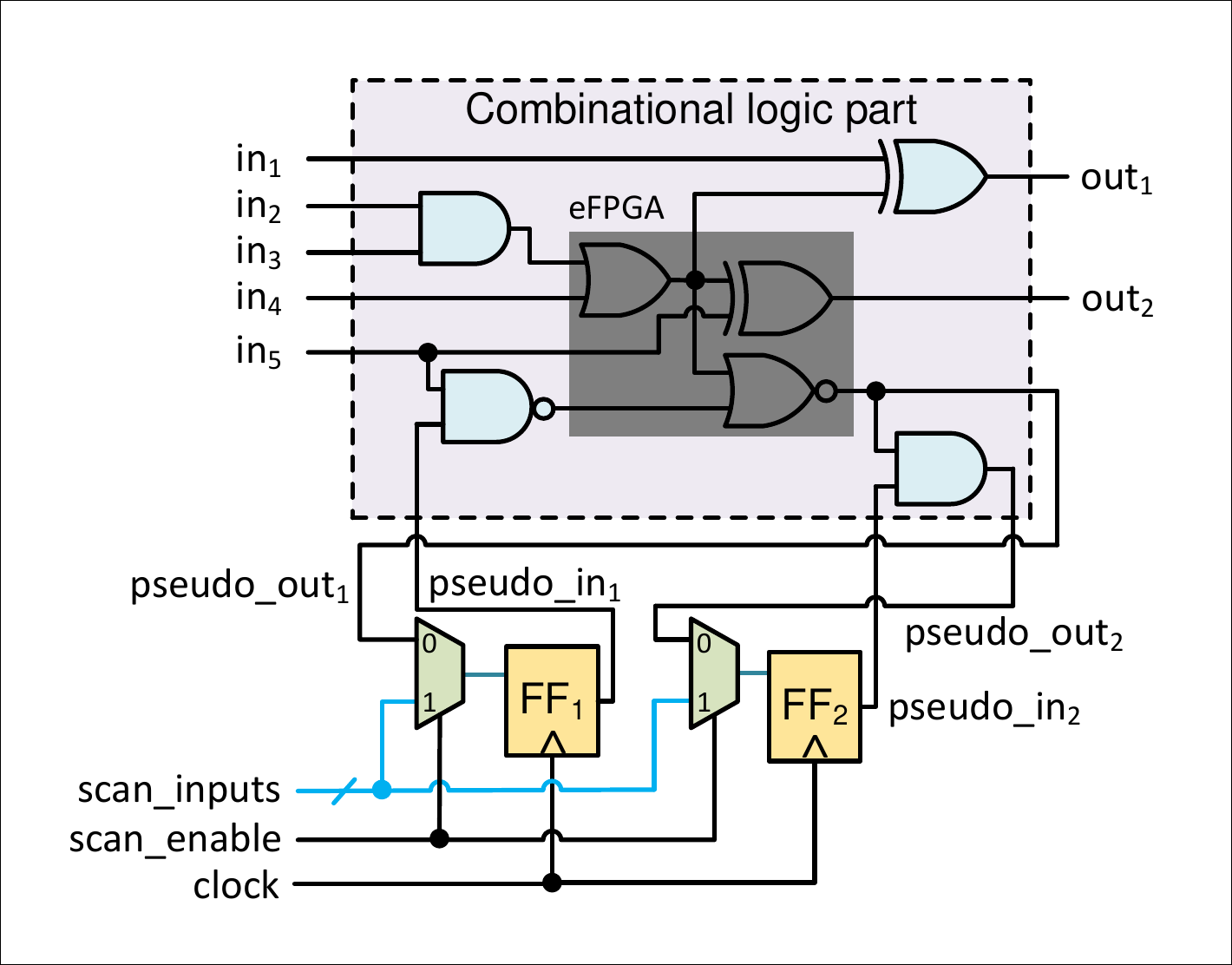}
    \caption{The 
    eFPGA redacted
    design with the scan chain.}
    \label{fig:seq_design_components}
\end{figure}

Similar to 
scan chains
in application-specific integrated circuits (ASICs), scan chains can be utilized in field-programmable gate arrays (FPGAs), 
as shown in Figure~\ref{fig:seq_design_components}.
Scan chains allow arbitrary test patterns to be loaded into flip-flops on an FPGA~\cite{tiwari2003scan,palchaudhuri2017redundant,renovell2001fpga}.
Note that testability is important for embedded FPGAs (eFPGAs)
as they are integrated with an ASIC design.

Usually, 
scan chains are not open to end-users or attackers. 
Techniques, such as flipped
scan~\cite{sengar2007secured}, XOR scan~\cite{agrawal2008scan}, double feedback XOR scan~\cite{banik2013improved}, and
sub-chains based scan~\cite{lee2005securing,lee2007securing,atobe2013secure,oya2014secure}, protect scan chains and restrict their access; however, they are broken by attacks, including Mukesh \textit{et al.}~\cite{agrawal2008scan},
    Subhadeep \textit{et al.}~\cite{banik2013improved},
    Banik \textit{et al.}~\cite{banik2014cryptanalysis},
    and Cui \textit{et al.}~\cite{cui2017current}.
Some techniques disable access to scan chains using AES algorithm~\cite{SecureScan}---however, such AES-based techniques are vulnerable to attacks, including DaRolt \textit{et al.}~\cite{DaRolt2011scanattackandcountermeaures} and Ali \textit{et al.}~\cite{ali2015novel}.
Some other scan-chain 
protections
were developed using the concept of logic locking,
such as
Encrypt Flip-Flop~\cite{karmakar2018encrypt}
and dynamic scan obfuscation (DynScan)~\cite{karmakar2019efficient}.
Yet, these scan-chain protections are vulnerable to attacks, 
such as ScanSAT~\cite{alrahis2019scansat} and DynUnlock~\cite{limaye2020dynunlock}.
In this case, the attacker is able to obtain unauthorized scan-chain access to perform \proposed~after the testing stage.
Currently, there are secure scan-chain techniques robust to all existing attacks, such as \textit{dishonest oracle} (\textit{DisORC})~\cite{limaye2020thwarting}. 
However, DisORC may not sufficiently protect a circuit from recovery by \proposed~because DisORC’s effectiveness is limited. For example, the Hamming distance (HD) between the correct output and the output on applying a random incorrect key is $11.30\%$ for b20 in DisORC~\cite{limaye2020thwarting}, but the ideal HD should be $50\%$ on average~\cite{rajendran2014regaining}.

\section{Scalably Expanding from Minterm to PI}
\label{sec:proposed_scable_expansion}

\begin{algorithm}[t]
\caption{Scalable PI expansion}
\label{alg:minterm2PI_scalability}

\small

\SetAlgoLined

\SetKwFunction{FMain}{\text{expand\_scalably}}
\SetKwFunction{FUpdateCube}{update\_cube}
\SetKwFunction{IterLimit}{get\_iter\_limit}
\SetKwFunction{GetHardDc}{get\_hard\_dc}

\SetKwProg{Fn}{Function}{:}{}

\SetKwProg{FORwithnoend}{for}{ do}{}
\SetKwProg{IFwithnoend}{if}{ then}{}
\SetKwProg{ELSEwithnoend}{else}{}{}

\KwIn{
{$\mathcal{O}$}, 
{$w$},
{$m_0$},
{$PIT_{pred}$},
{$p$},
{$p_0$}
}
\KwOut{$PI_{pred}$}

\Fn{ \GetHardDc {$m_0$, $PIT_{pred}$} }{
$hard\_dc := \varnothing$ \\
\FORwithnoend{$index \in \{1,\cdots, len(m_0)\}$}{

$hard\_dc\_flag := is\_dc\_always(index, PIT_{pred})$ \\ 
\IFwithnoend{$hard\_dc\_flag==True$}{
$hard\_dc := hard\_dc \bigcup \{ index \}$
}
}
}
{\bf return} $hard\_dc$ \\

\Fn{\IterLimit {$cube$,  $index$, $hard\_dc$, $p$, $p_0$ }}{
$num\_dc := cube.count(\text{``}\scalebox{1.35}{-}\text{''})$ \\
\IFwithnoend{$index \in hard\_dc $}{
$iter\_limit := min\{2^{num\_dc}, p_0 \}$ \\
}
\ELSEwithnoend{}{
$iter\_limit~:=~min\{2^{num\_dc}, p\times~num\_dc \}$ \\
}
}
{\bf return} $iter\_limit$ \\

\Fn{ \FMain {$\mathcal{O}$, $w$, $m_0$, $p$, $p_0$}}{

$hard\_dc :=$ \GetHardDc {$m_0$, $PIT_{pred}$} \\
$cube := m_0$  \\     
\FORwithnoend{$index \in \{1,2, \cdots, len(m_0) \}$ }{
$iter\_limit := $
\IterLimit{$cube$,  $index$, $hard\_dc$, $p$, $p_0$} \\

$cube:=$ \FUpdateCube($\mathcal{O}$, $w$, $cube$, $index$, $iter\_limit$) \hfill {$\rhd$ \texttt{update\_cube()} in Algorithm~\ref{alg:minterm2PI}}
}
$PI_{pred} := cube$ \\
}
{\bf {return}} $PI_{pred}$  \\

\end{algorithm}

Algorithm~\ref{alg:minterm2PI_scalability} provides a solution for scalable expansion from an ON-set minterm to a predicted prime implicant (PI) compared to Algorithm~\ref{alg:minterm2PI}.
In Algorithm~\ref{alg:minterm2PI_scalability}, The time taken in determining a PI is proportional to the number of don't cares. 
The non-effective inputs
are always don't cares in all the PIs.
We refer to these as hard don't care (\textit{hard\_dc}) bits.
To reduce the time taken in determining a potential don't care bit, the number of queries on 
the hard\_dc bit
is limited to a constant value 
$p_0$, where $p_0 < p \times num\_dc$ in most cases.
We consider $p_0$ as the constant limitation parameter in Algorithm~\ref{alg:minterm2PI_scalability}.
Suppose,
out of 
$n$ inputs,
there are $n_r$ effective 
inputs ($n_r<n$).
In the worst case, the complexity is $O(n_r^2)$, which is less than $O(n^2)$.
Thus, Algorithm~\ref{alg:minterm2PI_scalability} has higher efficiency than Algorithm~\ref{alg:minterm2PI} since $O(n_r^2)\ll O(n^2)$ when $n_r\ll n$.

\section{Analysis of Parameters in \proposed}
\label{sec:appendix_parameters_accuracy}

{\bf Time Limit $\mathbf{T}$.} 
With greater $T$,
\proposed~can achieve a higher prediction accuracy because
the total number of PIs increases.
{\bf Linear Limitation Parameter $\mathbf{p}$.}
In the worst case, the number of queries in each PI expansion is
$\sum\limits_{i=1}^{n}{p\times i} = p \times \frac{n^2+n}{2}$, where $n$ is the number of inputs. Thereby, in the worst case, the time complexity of each PI expansion is $O(n^2)$.
{\bf Constant Limitation Parameter $\mathbf{p_0}$.}
Suppose the PI expansion uses Algorithm~\ref{alg:minterm2PI_scalability}. Assume a predicted PI has $n_r$ effective inputs. In the worst case, the number of queries
in the PI expansion is
$\sum\limits_{i=1}^{n_r} p\times i + \sum\limits_{i=1}^{n-n_r} p_0 = p\times \frac{n_r^2 + n_r}{2} + p_0\times (n-n_r)$. 
Thus, the time complexity of the  scalable PI expansion is $O(n_r^2)$. When $n_r \ll n$ (most inputs are non-effective inputs), the acceleration of the PI expansion process (comparing Algorithm~\ref{alg:minterm2PI_scalability} to Algorithm~\ref{alg:minterm2PI}) is huge since $O(n_r^2) \ll O(n^2)$.
{\bf Distance Parameter $\mathbf{d_0}$.} In our simulations, we choose $d_0=2$ based on a case study of distance distribution on IBEX~\cite{woods2014ibex}, as described in Section~\ref{sec:distribution_hd}.
{\bf Convergence Parameter $\mathbf{p_{conv}}$.}
Let $c$ be the confidential probability of the following event:
{\it \proposed~consecutively visits ${p_{conv}}$ OFF-set minterms.}
Suppose $r^{ON}$ is the ratio of the ON-set minterms in the search space for the next ON-set minterm. 
Thus, $c=(1-r^{ON})^{p_{conv}}$.
Further, if we consider $p_{conv}$ as a fixed parameter, $r^{ON}=1-c^{\frac{1}{p_{conv}}}$.
In our simulations,
we choose $p_{conv}=50$. 
If this event happens, then
$10\% < c \le 100\%$ implies
$0\le r^{ON} < 4.5\%$.
Thus, after consecutively visiting OFF-set minterms for $p_{conv}=50$ times during the search for the next ON-set minterm, we conclude that 
ON-minterms are rare in the {current search space since $r^{ON}$ is negligibly small.

\section{FPGA Hardware Implementation}
\label{sec:appendix_demo}

\begin{figure}[b!]
    \centering
    \includegraphics[clip, trim=0.93cm 0.95cm 0.95cm 0.93cm,
    width=0.495\textwidth]{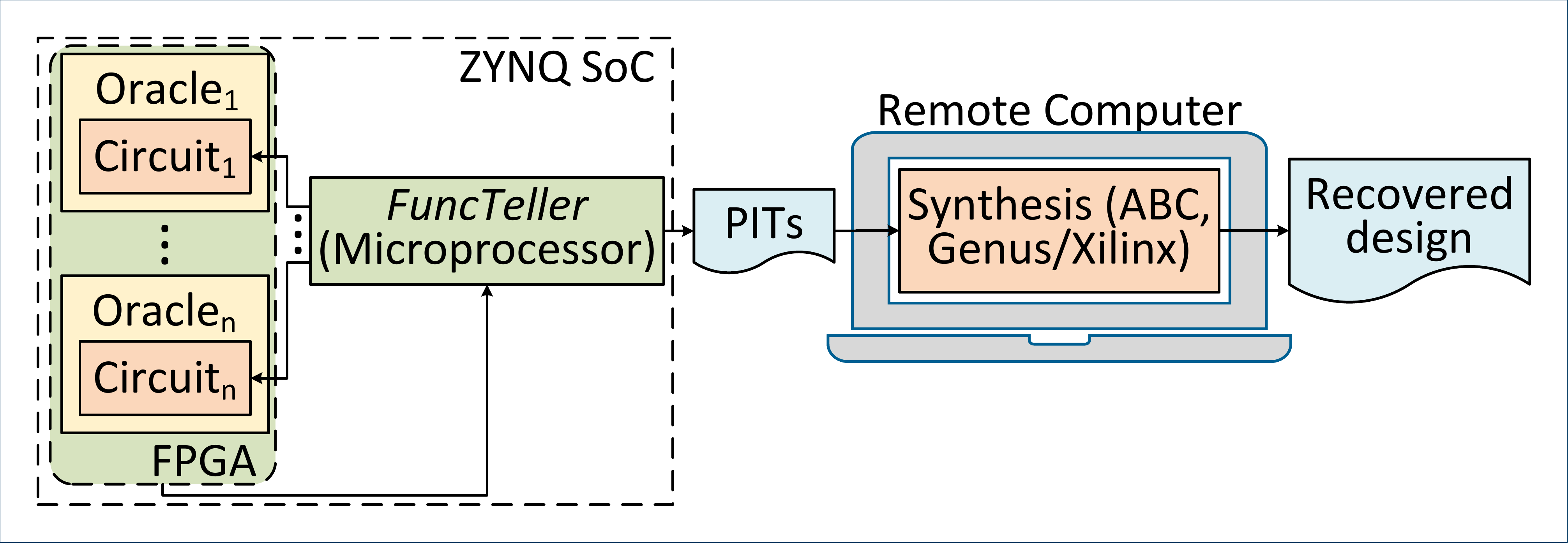}
    \caption{FPGA prototype setting.}
    \label{fig:fpga_implementation_done}
    \end{figure}

We construct an FPGA-based prototype to mimic an eFPGA by solving two challenges:
        (i)
        \textbf{High communication slack} 
        between \proposed~and oracle affects the attack time significantly. To reduce this communication slack, we implement \proposed~and oracle on 
        the same
        SoC platform.
        (ii)
        \textbf{Lack of parallelization:} To match the attack time of the software-based results of \proposed, we implement multiple oracles on the same FPGA. The number of oracles 
        on the FPGA 
        is
        limited by resource constraints.

    \noindent
    \textbf{Implementation.} We attack the circuits in Table~\ref{tab:scale_and_results_benchmark_and_processor} using the FPGA prototype shown in Figure~\ref{fig:fpga_implementation_done}, using the PYNQ-Z2 board.
    The PYNQ-Z2 board contains a Zynq-7000 series SoC platform containing both an FPGA (performing oracle) and a processor (running \proposed). 
    We use the Xilinx AXI4 interface~\cite{DisableBitstreamXilinx} to reduce communication slack. Further, for larger designs, we repeat this process for multiple boards. 
    Finally, all PITs are sent to a remote computer. 

    \noindent
    \textbf{Results and Analysis.} We empirically observe that the attack needs approximately twice as long to recover the design with the same accuracy compared to the simulation-based version. These results on selected circuits are shown in Table~\ref{tab:results_fpga_done}. While the software simulation runs on an Intel Xeon processor running at 2.6GHz, our hardware implementation is on a 650MHz dual-core Cortex-A9 processor. This leads to a degradation in performance. Thus, the time taken per query is approximately $2\times$ the time taken per query in the simulation-based setting, as there is a software overhead while generating new queries. In this case, we are limited due to SoCs available.
    Note that the three
    circuits (b20, IBEX, and GPS)
    in Table~\ref{tab:scale_and_results_benchmark_and_processor} are not included in Table~\ref{tab:results_fpga_done}. Due to the FPGA resource constraints, the extraction of these designs was not complete. To further explain this, we present a scalability analysis for these circuits.

    \noindent
    \textbf{Scalability Analysis.} On the one hand, the circuits (IBEX, GPS, and b20) excluded from Table~\ref{tab:results_fpga_done} have a large number of input/output ports; on the other hand, the interfaces available on the SoC platform are limited. As a result, the number of oracles that can be implemented on the SoC 
    is
    limited, and consequently, \proposed~cannot
    be completed.
    For instance, we can only implement $5$ IBEX oracles on a single SoC. This can be overcome by parallelizing the attack on multiple SoCs to recover the entire design. Due to the lack of such resources, we are unable to parallelize to the extent possible in software (50 outputs in parallel). Next, we estimate \proposed's performance if additional SoCs were available.
    
Table~\ref{tab:estiamte_fpga_done} shows the estimated time needed to match the simulation-based accuracy if the number of oracles is $50$. Recall that the number of oracle implementations on a single SoC is limited by the number of AXI4 interfaces. Thus, more SoCs are needed to parallelize the attack to at least $50$ oracles. We derive the time estimate in Table~\ref{tab:estiamte_fpga_done} based on empirical observations and validate these observations based on tests for individual outputs. As stated previously, \proposed~needs approximately twice as long to recover the design if the number of oracles is the same. This is due to the communication slack being 
$2\times$
higher than the software-based implementation of \proposed. Notably, the communication slack is only 
$2\times$
higher while the Cortex-A9 processor is approximately 
$3\times$
slower compared to the Intel Xeon processor used for simulation. Based on this analysis, we can conclude that powerful SoCs which have a large number of AXI interfaces and powerful onboard CPUs can effectively recover large-scale redacted designs as \proposed~can scale to larger designs depending 
on
the hardware platforms used.

    \begin{table}[t]
    \centering
    \caption{The results of \proposed~on the 
    FPGA prototype.}
    \resizebox{0.34\textwidth}{!}{
    \begin{tabular}{|c|c|c|c|}
    \hline
    \textbf{Circuit}
    & \textbf{\begin{tabular}[c]{@{}c@{}}
    \# oracles
    \end{tabular}} & \textbf{\begin{tabular}[c]{@{}c@{}}Attack time \\ (hours)\end{tabular}} & \textbf{\begin{tabular}[c]{@{}c@{}}Accuracy \\ (\%)\end{tabular}} \\ \hline \hline
    c432               & 7                                                                                    & 0.16                                                                & 99.26                                                             \\ \hline
    c880               & 26                                                                                   & 1.96                                                                & 95.23                                                             \\ \hline
    c1355              & 32                                                                                   & 2.34                                                                & 52.33                                                             \\ \hline
    c1908              & 25                                                                                   & 2.21                                                                & 79.43                                                             \\ \hline
    c7552              & 46                                                                                   & 2.53                                                                & 88.02                                                             \\ \hline
    b14                & 107                                                                                  & 1.97                                                                & 90.34                                                             \\ \hline
    MIPS               & 75                                                                                   & 2.63                                                                & 93.24                                                             \\ \hline
    \end{tabular}
    }
    \label{tab:results_fpga_done}
    \end{table}

    \begin{table}[t]
    \centering
    \caption{Estimated time and resources for \proposed~to match the simulation-based accuracy on 
    FPGAs.}
    \resizebox{0.38\textwidth}{!}{
    \begin{tabular}{|c|c|c|c|}
    \hline
    \textbf{Circuit} & \textbf{\begin{tabular}[c]{@{}c@{}}
    {Max. \# oracles}\\ on each FPGA
    \end{tabular}} & \textbf{\begin{tabular}[c]{@{}c@{}}
    \# FPGAs\\ required
    \end{tabular}} & \textbf{\begin{tabular}[c]{@{}c@{}}Estimated\\ time (hours)\end{tabular}} \\ \hline \hline
    b20                & 15                                                                                   & 4                                                                         & 20                                                                                                     \\ \hline
    IBEX               & 5                                                                                    & 10                                                                        & 144                                                                                                    \\ \hline
    GPS                & 1                                                                                    & 50                                                                        & 88                                                                                                     \\ \hline
    \end{tabular}
    }
    \label{tab:estiamte_fpga_done}
    \end{table}

\begin{table}[t!]
    \centering
    \caption{Comparison of performance between \proposed~and Chen \textit{et al.}~\cite{chen2020circuit}
    on CSAW 2021 competition circuits~\cite{csaw2021}.}
    \resizebox{0.485\textwidth}{!}{
    \begin{tabular}{|c|c|c|c|c|c|}
    \hline
    \multicolumn{2}{|c|}{\multirow{2}{*}{\bf Circuit}} &\multirow{2}{*}{\bf \# inputs}&\multirow{2}{*}{\bf \# outputs} & \multicolumn{2}{c|}{\bf Accuracy (\%)} \\ \cline{5-6}
    \multicolumn{2}{|c|}{}&&    &{\bf {\bf Chen \textit{et al.}~\cite{chen2020circuit}}} &{\bf \proposed}  \\ \hline\hline
\multirow{28}{*}{\rotatebox{90}{\bf MAIN}}  
&set1\_1	
& $	94	$&$	90	$&$83.11 $  &$100.00$   \\ \cline{2-6}
&set1\_2	
& $	98	$&$	94	$&$83.33$   &$100.00$  \\ \cline{2-6}
&set1\_3    
& $	102	$&$	98	$&$78.40$   &$100.00$  \\ \cline{2-6}
&set1\_4	
& $	106	$&$	102	$&$78.29$   &$100.00$   \\ \cline{2-6}
&set1\_5	
& $	110	$&$	106	$&$76.73$   &$99.68$  \\ \cline{2-6}
&set1\_6	
& $	114	$&$	110	$&$75.78$   &$99.68$   \\ \cline{2-6}
&set1\_7	
& $	118	$&$	114	$&$78.19$   &$99.71$   \\ \cline{2-6}
&set2\_1	
& $	94	$&$	90	$&$84.26$   &$100.00$    \\ \cline{2-6}
&set2\_4	
& $	106	$&$	102	$&$76.34$   &$99.81$  \\ \cline{2-6}
&set2\_5	
& $	110	$&$	106	$&$75.82$   &$98.63$  \\ \cline{2-6}
&set2\_6	
& $	114	$&$	110	$&$76.71$   &$99.41$  \\ \cline{2-6}
&set2\_7	
& $	118	$&$	114	$&$78.47$   &$98.51$  \\ \cline{2-6}
&set3\_1	
& $	94	$&$	90	$&$83.34$   &$99.82$   \\ \cline{2-6}
&set3\_2	
& $	98	$&$	94	$&$82.92$   &$99.70$   \\ \cline{2-6}
&set3\_3	
& $	102	$&$	98	$&$77.57$   &$99.79$   \\ \cline{2-6}
&set3\_4
& $	106	$&$	102	$&$76.39$   &$98.90$   \\ \cline{2-6}
&set3\_5	
& $	110	$&$	106	$&$75.27$   &$99.17$   \\ \cline{2-6}
&set3\_6	
& $	114	$&$	110	$&$76.70$   &$99.08$   \\ \cline{2-6}
&set3\_7	
& $	118	$&$	114	$&$78.84$   &$98.80$   \\ \cline{2-6}
&set4\_1	
& $	94	$&$	90	$&$82.24$   &$100.00$    \\ \cline{2-6}
&set4\_2	
& $	98	$&$	94	$&$81.97$   &$100.00$   \\ \cline{2-6}
&set4\_3	
& $	50	$&$	46	$&$92.36$   &$100.00$   \\ \cline{2-6}
&set4\_4	
& $	186	$&$	182	$&$65.30$   &$99.42$   \\ \cline{2-6}
&set4\_5	
& $	110	$&$	106	$&$76.75$   &$99.44$   \\ \cline{2-6}
&set4\_6	
& $	114	$&$	110	$&$75.82$   &$99.47$   \\ \cline{2-6}
&set4\_7	
& $	118	$&$	114	$&$77.99$   &$99.20$   \\ \hline
\multirow{7}{*}{\rotatebox{90}{\bf BONUS} }
&	set1	
& $	198	$&$	194	$&$63.15$   &$95.84$  \\ \cline{2-6}
&	set2	
& $	214	$&$	210	$&$62.71$   &$96.13$  \\ \cline{2-6}
&	set4	
& $	246	$&$	242	$&$66.42$   &$96.23$  \\ \cline{2-6}
&	set5	
& $	262	$&$	258	$&$67.79$   &$96.74$  \\ \cline{2-6}
&	set6	
& $	278	$&$	274	$&$67.21$   &$96.65$  \\ \cline{2-6}
&	set7
& $	294	$&$	290	$&$67.83$   &$96.62$  \\ \hline
    \end{tabular}
     }
    \label{tab:csaw_circuits_attack_results}
\end{table}

\section{Results on CSAW Benchmarks}
\label{sec:appendix_csaw}
Table~\ref{tab:csaw_circuits_attack_results} shows \proposed's performance on the FPGA-based circuits from
Cybersecurity Awareness Worldwide (CSAW)
2021 logic locking event.
For all circuits across the main benchmark sets, 
\proposed~has an accuracy of $>90\%$.

\end{document}